\documentclass[journal]{IEEEtran}
\usepackage{caption}
\usepackage{amssymb}
\usepackage[cmex10]{amsmath}
\usepackage{stfloats}
\usepackage{graphicx}
\usepackage{subfigure}
\usepackage{tabularx}
\usepackage{booktabs} 
\usepackage{threeparttable}
\usepackage{epsfig,epsf,color,balance,cite}
\usepackage{verbatim}
\usepackage{url}
\usepackage{bm}

\usepackage{algorithm}
\usepackage{algorithmic}
\usepackage{multirow}
\usepackage{makecell}
\usepackage{longtable}

\usepackage{graphicx}
\usepackage{multirow}
\usepackage{multicol}

\usepackage{supertabular}
\usepackage{longtable}

\usepackage{color}
\definecolor{myc1}{rgb}{0,0,1}

\usepackage{tikz}
\usetikzlibrary{positioning, fit, calc}   
\usepackage{xcolor}  
\tikzset{block/.style={draw, thick, text width=1.35cm , minimum height=0.5cm, align=center},   
block1/.style={draw, thick, text width=1.2cm , minimum height=0.4cm, align=center},
line/.style={-latex}     
}  
\usepackage{pgfplots}
\pgfplotsset{compat=newest}
\usetikzlibrary{plotmarks}
\usetikzlibrary{arrows.meta}
\usepgfplotslibrary{patchplots}

\begin{document}
 
\title{A New Pathway to Integrated Learning and Communication (ILAC): Large AI Model and  Hyperdimensional Computing for Communication} 
\author{
%
{Wei Xu, \emph{IEEE Fellow}, Zhaohui Yang, Derrick Wing Kwan Ng, \emph{IEEE Fellow}, Robert Schober, \emph{IEEE Fellow}, \\ H. Vincent Poor, \emph{IEEE Fellow}, Zhaoyang Zhang, and Xiaohu You, \emph{IEEE Fellow}
}
\\
\textit{(Invited Paper)}
\vspace{-5mm}
\thanks{ Wei Xu and Xiaohu You are with the National Mobile
 Communications Research Laboratory (NCRL), Southeast University, Nanjing
 210096, China, and also with Purple Mountain Laboratories, Nanjing 211111, China (e-mail: wxu@seu.edu.cn, xhyu@seu.edu.cn).}
\thanks{Zhaohui Yang and Zhaoyang Zhang are with Zhejiang Laboratory, Hangzhou 311121, China,
 and also with the College of Information Science and Electronic Engineering, Zhejiang University, Hangzhou, Zhejiang 310027, China (e-mail:  yang\_zhaohui@zju.edu.cn, ning\_ming@zju.edu.cn).}

 \thanks{Derrick Wing Kwan Ng is with the School of Electrical Engineering and
 Telecommunications, The University of New South Wales, Sydney, NSW
 2052, Australia (e-mail: w.k.ng@unsw.edu.au).}
 \thanks{Robert Schober is with the Institute for Digital Communications
 (IDC), Friedrich-Alexander University Erlangen-Nuremberg, 91054 Erlangen,
 Germany (e-mail: robert.schober@fau.de).}
 \thanks{H. Vincent Poor is with the Department of Electrical and Computer
 Engineering, Princeton University, Princeton, NJ 08544 USA (e-mail:
 poor@princeton.edu).}
}
\maketitle

\begin{abstract} 

The rapid evolution of forthcoming sixth-generation (6G) wireless networks necessitates the seamless integration of artificial intelligence (AI) with wireless communications to support emerging intelligent applications that demand both efficient communication and robust learning performance. This dual requirement calls for a unified framework of integrated learning and communication (ILAC), where AI enhances communication through intelligent signal processing and adaptive resource management, while wireless networks support AI model deployment by enabling efficient and reliable data exchanges. 
However, achieving this integration presents significant challenges in practice. Communication constraints, such as limited bandwidth and fluctuating channels, hinder learning accuracy and convergence. Simultaneously, AI-driven learning dynamics, including model updates and task-driven inference, introduce excessive burdens on communication systems, necessitating flexible, context-aware transmission strategies. 
This paper provides a comprehensive overview of ILAC system design and optimization strategies. 
We establish corresponding foundational principles, covering system architectures and presenting a unified optimization formulation that closely links learning performance with communication efficiency. We then review recent advancements in ILAC from the strategic perspectives of model and data distribution, computational complexity, and communication overhead. 
Despite considerable progress, existing ILAC approaches still suffer from high communication overhead, unstable convergence, and scalability challenges. To address these issues, we propose an enhanced ILAC framework incorporating large AI models and hyperdimensional computing (HDC). In particular, utilizing large AI models improves generalization capabilities under dynamic task and network conditions, while HDC provides lightweight, high-dimensional representations that reduce both communication and learning costs. 
Finally, we present a case study on a cost-to-performance optimization problem, where task assignments, model size selection, bandwidth allocation, and transmission power control are jointly optimized, considering computational cost, communication efficiency, and inference accuracy. Leveraging the Dinkelbach and alternating optimization algorithms, we offer a practical and effective solution to achieve an optimal balance between learning performance and communication constraints.

\end{abstract}


\IEEEpeerreviewmaketitle

\section{Introduction}\label{Introduction}

\subsection{Motivation of ILAC}
\IEEEPARstart{T}{he} evolution toward sixth-generation (6G) wireless networks has introduced unprecedented demands that surpass the capabilities of state-of-the-art communication systems~\cite{8869705}. With the rapid proliferation of applications, such as autonomous vehicles, smart cities, and virtual reality, modern networks must be exceptionally adaptive to accommodate dynamic environments and heterogeneous service requirements. These advanced applications typically require real-time decision-making and context-aware resource allocation capabilities, which traditional communication systems often struggle to deliver~\cite{gmml,ZHAO2024107055,10988619}. In fact, conventional networks depend heavily on predefined protocols and static configurations, rendering them inadequate for handling fluctuating network conditions and diverse application needs. Consequently, there is a growing imperative to combine artificial intelligence (AI) techniques with communication systems, which is a key element of the future vision of 6G. Specifically, AI-driven approaches enable efficient real-time decision-making and advanced pattern recognition, thereby facilitating the optimization of network performance and resource utilization to satisfy the demands of next-generation wireless technologies~\cite{wang2024multiloc,ncps_beamforming,zhu2024robust,wu2023technique}. 
Nevertheless, the application of  AI models to wireless networks also presents significant challenges, especially for large-scale AI models. In practice, these models typically require extensive computational resources, which traditional centralized computing systems are often unable to support efficiently.

To overcome this limitation, distributed computing has emerged as a viable and promising solution~\cite{9210812}. By leveraging intelligent connectivity, distributed computing enables the dynamic allocation of computing power and data collection across the network, thereby enhancing model scalability and facilitating efficient deployment of AI in diverse and resource-constrained environments.

Nevertheless, despite these benefits, the dual requirements of high-performance communication and computation-intensive AI tasks present a fundamental conflict due to limited resources, including spectrum and computational capacity~\cite{10915662,wgan,zhao2025energy,xu2025data}, available in 6G networks. Furthermore, emerging applications place simultaneous pressures on both communication and AI infrastructure, requiring not only high-bandwidth, low-latency connectivity but also substantial computing power for real-time analytics, AI inference, and deep learning (DL).

Traditional communication systems address communication and learning tasks independently,  resulting in suboptimal designs that inadequately allocate resources across both domains. In practice, this fragmented optimization leads to resource contention and inefficient utilization. Moreover, the inherent static assumptions in traditional systems severely limit their adaptability to dynamic network conditions, such as fluctuating data distributions, varying traffic loads, and evolving quality of service (QoS) requirements ~\cite{6926773,yao2023imperfect}. This mismatch with real-time demands deteriorates system performance and reduces flexibility in handling emerging applications.
Thus, a unified approach that jointly optimizes learning and communication processes is crucial to effectively address these challenges. 

Integrated learning and communication (ILAC) offers a unified framework that optimizes both communication and learning objectives simultaneously, ensuring efficient resource utilization, enhanced adaptability to dynamic environments, and superior performance in resource-constrained 6G networks, e.g., \cite{10504775,10411061,duan2019state,7225113,8572766}.
In contrast, traditional AI models, while effective for specific tasks, face considerable limitations in next-generation networks. Specifically, these models are typically optimized for predefined scenarios and thus require frequent retraining and redeployment to adapt to new environments or tasks—processes that are both time and resource-intensive. This rigidity becomes particularly problematic in dynamic 6G networks, where diverse applications demand continuous adaptation  \cite{shi2022intelligent}. To address these limitations, the ILAC paradigm integrates large AI models through their inherent generalizability and multi-tasking capabilities. Trained on massive datasets with large-scale model parameters, these models can handle diverse tasks in multiple scenarios without the need for task-specific optimization \cite{10579546}. Indeed, by enabling a single large AI model to perform multiple functions, ILAC significantly reduces the operational overhead associated with deploying and maintaining specialized models, making it especially suitable for the complex and evolving demands of 6G networks and beyond.

\begin{figure*}[t]
\centering
\includegraphics[width=.95\linewidth]{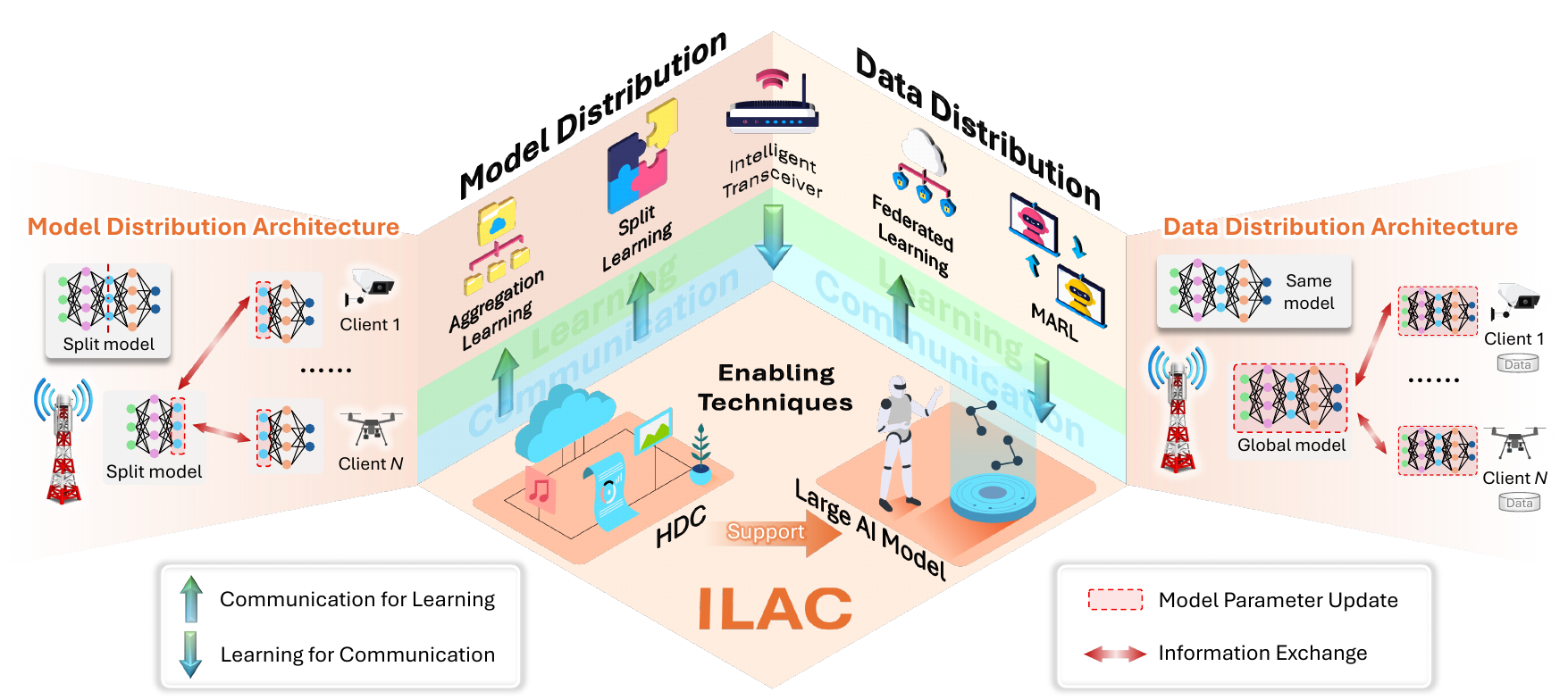}
\caption{Illustration of ILAC.}
\label{fg:s1:ilac}
\end{figure*}

To facilitate their versatility, large AI models typically comprise billions of parameters, imposing  unprecedented demands on both computational and communication resources within ILAC networks. To address these challenges, hyperdimensional computing (HDC) has emerged as a potential solution \cite{Kleyko2022SurveyHyperdimensional,Kleyko2023SurveyHyperdimensional}. Inspired by the mathematical properties of high-dimensional spaces, HDC provides a lightweight yet robust computational framework that efficiently represents and processes information by leveraging hyperdimensional vectors. Unlike traditional numerical computation methods, HDC encodes data into high-dimensional spaces, where information is distributed across numerous dimensions. This distributed characteristic not only enhances robustness against noise and errors but also facilitates highly efficient parallel processing. Moreover, the inherent redundancy in high-dimensional representations ensures that even if partial information is lost across some dimensions, the overall integrity of the encoded data remains intact. Furthermore, due to their element-wise independence, hyperdimensional operations naturally lend themselves to parallel computing architectures.

In practice, integrating HDC into ILAC offers a promising approach to alleviate the computational burden associated with large AI models. In particular, HDC simplifies learning operations by reducing complex computations to basic mathematical manipulations on high-dimensional vectors, such as additions and multiplications. Moreover, its capability to rapidly adapt to diverse tasks without requiring extensive retraining makes it particularly well-suited for addressing the dynamic and heterogeneous service requirements inherent to 6G networks.

In summary, ILAC represents a potentially transformative paradigm shift for communication networks, evolving from traditional data transmission systems to intelligent, adaptive, and context-aware infrastructures. By overcoming the inefficiencies of legacy systems and integrating advanced technologies such as large AI models and HDC, ILAC is poised to revolutionize how communication networks support next-generation intelligent applications. This integrated approach empowers future 6G networks to fulfill the stringent demands for intelligent and efficient communication, enabling groundbreaking advancements across diverse industries and applications~\cite{10550151,10552192,9410457}.

\subsection{Unified ILAC Framework}


As illustrated in Fig.~\ref{fg:s1:ilac}, ILAC can be generally categorized into two primary architectures: model distribution and data distribution, which together facilitate a tight synergy between learning and communication functionalities.

Specifically, the model distribution architecture emphasizes the efficient deployment, continuous adaptation, and efficient collaborative sharing of AI learning tasks. In practice, key approaches include split learning~(SL), which distributes model training across devices and servers \cite{lin2024split}; aggregation learning (AL), which amalgamates models from multiple devices \cite{duan2019state}; and intelligent transceivers, which jointly optimize model parameter updates while embedding learning capabilities into communication hardware. These techniques enable scalable and adaptive model utilization, especially in the highly distributed environments anticipated for future wireless networks.

The data distribution architecture effectively coordinates data handling and dissemination across the network to improve learning and communication performance. Core techniques such as federated learning (FL), which enables decentralized training while preserving data privacy by avoiding  the sharing of raw data \cite{yao2023gomore,shi2024empowering}; and multi-agent reinforcement learning (MARL), which allows multiple agents to collaboratively develop optimized strategies by exploiting local data \cite{foerster2016learning}, play a critical role. Additionally, through adaptive localized learning, intelligent transceivers can dynamically adjust data transmission and reception based on  network conditions and user requirements.

Beyond the architectures of model and data distribution, ILAC also incorporates a suite of advanced techniques to further enhance its capabilities. In particular, large-scale AI models provide robust generalization across diverse application scenarios, while HDC alleviates the computational burden associated with large AI models, thereby facilitating their efficient deployment and execution in resource-constrained environments.

\subsection{Focus and Structure}

\begin{table*}[t]
\renewcommand\arraystretch{1.3} 
\centering
\caption{Comparison of Key Surveys and Our Contribution}
\label{Summary}
\begin{tabular}{|c|c|m{3.7in}|m{2.2in}|}
\hline
\textbf{Year} & \textbf{Ref.} & \textbf{Core Contribution} & \textbf{Technical Focus} \\ \hline
2017 & \cite{7879258} & Pioneering survey on mobile edge computing (MEC) architectures and computation offloading strategies & MEC infrastructure; computation offloading \\ \hline
2018 & \cite{8110603} & Foundational taxonomy for contextual intelligence and data analytics in IoT& Context-aware computing in IoT ecosystems \\ \hline
2019 & \cite{8743390} & Systematic review of ML techniques in PHY/MAC layer optimization& ML methods for wireless resource management \\ \hline
2019 & \cite{8884164} & Edge intelligence framework design tailored to IoV applications& Edge caching/computing for vehicular networks \\ \hline
2020 & \cite{9060868} & Investigation of FL deployment challenges in resource-constrained edge networks & FL communication-cost tradeoffs \\ \hline
2021 & \cite{9352033} & Development of a three-tier classification framework for FL in networking & FL privacy-security co-design \\ \hline
2021 & \cite{9357490} & Formulation of cross-layer optimization principles for distributed learning & Communication-efficient ML parameter tuning \\ \hline
2021 & \cite{9519636} & Unified model for resource scheduling in edge computing& Computation offloading and provisioning \\ \hline
2021 & \cite{9562559} & Exploration of emerging paradigms in wireless distributed learning & FL deployment under wireless constraints \\ \hline
2022 & \cite{9475501} & FL adaptation strategies for heterogeneous IoT devices & Addressing device heterogeneity in FL systems \\ \hline
2022 & \cite{9606720} & Service-driven edge AI architecture for 6G networks & Sensing-communication-computation integration \\ \hline
2023 & \cite{10024766} & Review of EL techniques for B5G semantic communication & Semantic-aware network optimization \\ \hline
\textbf{2025} & \textbf{This paper} & \textbf{1)ILAC framework with a dual model/data distribution taxonomy;} \textbf{2) Integration of large AI models and HDC for ILAC} & \textbf{FL, MARL, intelligent transceiver, SL, AL; large AI models, HDC} \\ \hline
\end{tabular}
\end{table*}


\begin{figure}[t]
\centering
\includegraphics[width=\linewidth]{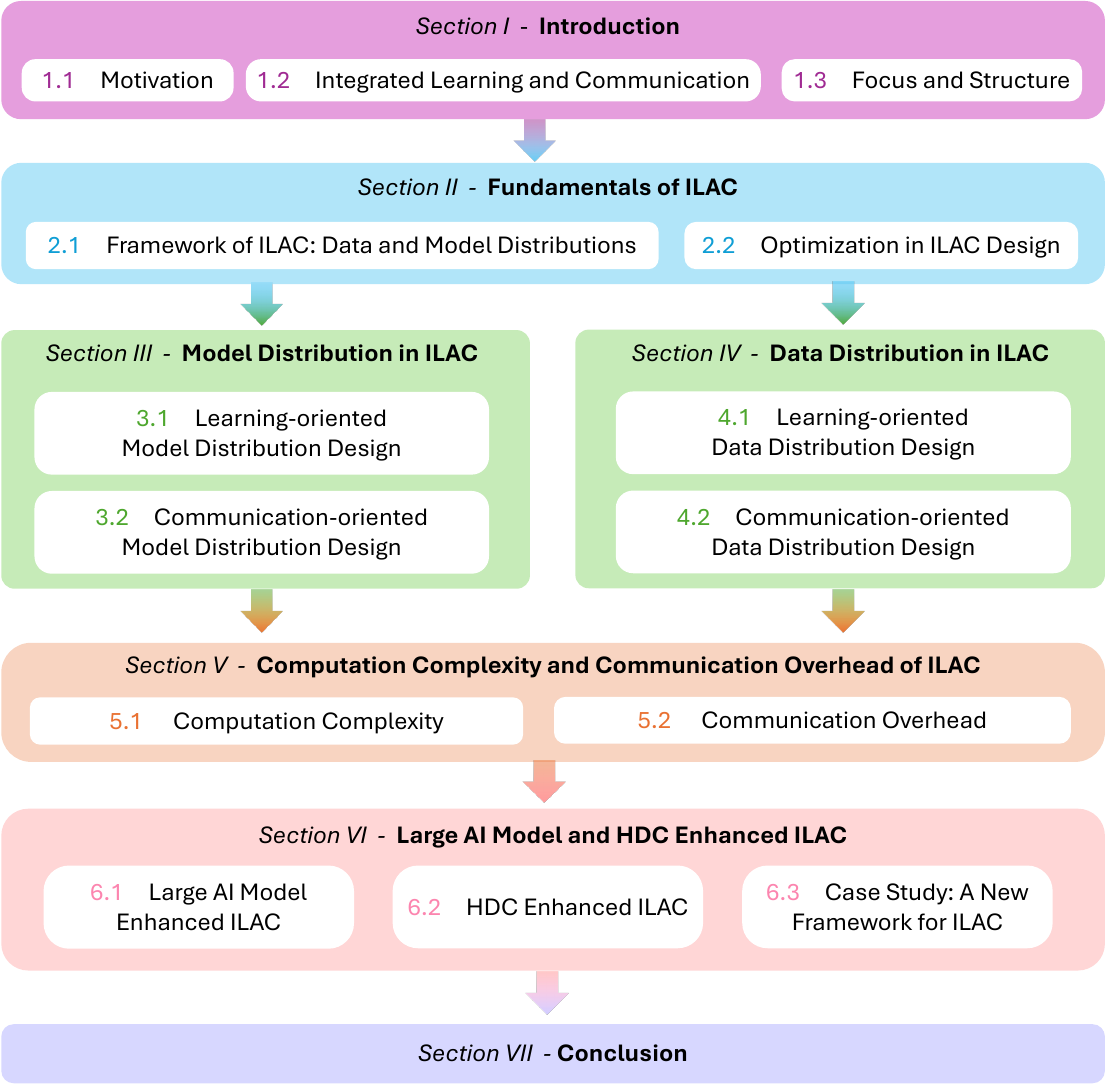}
\caption{Structure of the paper: An overview of the main topics.}
\label{fg:s1:ps}
\end{figure}

There have been several surveys exploring the integration of learning and communication in various forms. For instance, the authors in \cite{7879258,8884164,9519636} provided comprehensive reviews on edge computing. Specifically, \cite{7879258} focused on computation offloading, \cite{8884164} emphasized intelligent task enhancement for the Internet-of-Vehicles (IoV), and \cite{9519636} concentrated on resource scheduling in edge computing. Furthermore, techniques for edge learning (EL) in distributed wireless networks were reviewed in \cite{8110603,8743390,9606720,10024766}. In particular, \cite{8110603} examined a context-aware computing scheme for the Internet-of-Things (IoT), \cite{8743390} and \cite{9606720} covered machine learning (ML) and the integration of edge AI applications in wireless networks, while \cite{10024766} focused on distributed signal processing approaches in beyond 5G (B5G) networks. Regarding distributed learning, \cite{9357490} and \cite{9562559} provided comprehensive overviews, with the former focusing on improving communication efficiency and the latter on efficient deployment by considering wireless constraints and privacy. Furthermore, the studies in  \cite{9060868,9352033}, and \cite{9475501} highlighted FL integration, providing in-depth tutorials covering communication efficiency, resource management, and privacy challenges.
Table \ref{Summary} summarizes the aforementioned surveys, detailing their primary contributions and areas of focus.

Unlike existing surveys, this work introduces a novel taxonomy that categorizes ILAC into two primary paradigms: model distribution and data distribution. First, we provide a formal definition of ILAC as a unified framework that seamlessly integrates precise learning and efficient communication in wireless networks. Then, we discuss the design and optimization of ILAC in depth, highlighting key algorithms and applications. Additionally, we explore the incorporation of large AI models and HDC into ILAC, emphasizing how these techniques significantly enhance system performance.

As outlined in Fig.~\ref{fg:s1:ps}, the paper is organized as follows. Section \ref{Sec:fi} introduces the fundamentals of ILAC, beginning with an overview of the general framework, followed by discussion of the architectures for model and data distributions, as well as the optimization design. In Sections \ref{Sec:mdai} and \ref{Sec:ddai}, we delve into model distribution and data distribution in ILAC, respectively, each addressing both learning-oriented design and communication-oriented design in both cases. The computational complexity and communication overhead associated with ILAC are discussed in Section~\ref{Sec:ccco}. Section~\ref{Sec:lhc} introduces an enhanced ILAC framework that exploits large-scale AI models and HDC. Finally, conclusions are drawn in Section \ref{Sec:c}.
For the reader's convenience, a list of key abbreviations used is provided in Table \ref{KeyAcronym}.

\begin{table}[t]
\centering
\caption{List of Key ABBREVIATIONS}
\label{KeyAcronym}
\setlength{\tabcolsep}{5.2mm}
\begin{tabular}{|c|l|}
\hline
\textbf{Acronym}   & \textbf{Description}                     \\ \hline
6G           & The Sixth-Generation Mobile Networks   \\ \hline
AI           & Artificial Intelligence            \\ \hline

CSI             &Channel State Information      \\
\hline

DL           & Deep Learning                 \\ \hline
DNN          & Deep Neural Network           \\ \hline
DP           & Differential Privacy          \\ \hline
DRL          & Deep Reinforcement Learning   \\ \hline
EL           & Edge Learning                \\ \hline
FL           & Federated Learning            \\ \hline
HDC          & Hyperdimensional Computing \\ \hline

ILAC         & Integrated Learning and Communication    \\ \hline
IoT          & Internet-of-Things          \\ \hline
IoV          & Internet-of-Vehicles          \\ \hline
RIS          & Reconfigurable Intelligent Surface  \\ \hline 


LLM         & Large Language Model       \\ \hline
LoRA         & Low-Rank Adaptation     \\ \hline
MARL         & Multi-Agent Reinforcement Learning            \\ \hline
MEC          & Mobile Edge Computing         \\ \hline
ML           & Machine Learning              \\ \hline


QoS          & Quality of Service            \\ \hline


SFL          & Split Federated Learning  \\ \hline
SGD          & Stochastic Gradient Descent   \\ \hline
SL           & Split Learning               \\ \hline
UAV          & Unmanned Aerial Vehicle       \\ \hline

\end{tabular}
\end{table}

\section{Fundamentals of ILAC}\label{Sec:fi}
In this section, we focus on the fundamental aspects of ILAC, introducing the core design principles of a unified ILAC framework and its essential components specifically tailored for 6G networks. Based on these, two foundational architectures of ILAC are detailed, i.e., model distribution and data distribution. We formulate the ILAC design as a generalized optimization problem, which optimizes the dual functional objectives of learning and communication by jointly optimizing computing resources, learning algorithms, and transmission strategies to achieve performance objectives subject to various practical resource constraints.

\subsection{Framework of ILAC: Model and Data Distributions}\label{2A}
In this subsection, we first introduce the unified ILAC framework, outlining its core design principles and essential components. We then refine it from the perspective of model distribution and data distribution. Fig.~\ref{ILAC system} illustrates the proposed unified ILAC framework, where learning and communication not only coexist but also synergistically enhance each other's performance. The following discussion presents the primary design principles and components of the ILAC framework. Specially, the ILAC design follows three fundamental principles.
\begin{itemize}
\item \textbf{Communication efficiency}: Communication efficiency constitutes a core design objective of the ILAC framework, aimed at minimizing communication costs while ensuring reliable information exchange. To achieve this, ILAC prioritizes two key aspects: resource optimization and adaptive transmission strategies. First, it reduces communication overhead, latency, and bandwidth consumption by eliminating redundant transmissions and carefully refining the volume and frequency of data exchange. Second, it enhances adaptability by dynamically adjusting transmission parameters, such as data rates and coding schemes, in response to real-time channel state information (CSI), evolving user demands, and network dynamics. These capabilities allow ILAC to efficiently support ultra-low-latency, high-density applications, such as intelligent healthcare and industrial IoT deployments.


\begin{figure}[t]
\centering
\includegraphics[width=1\linewidth]{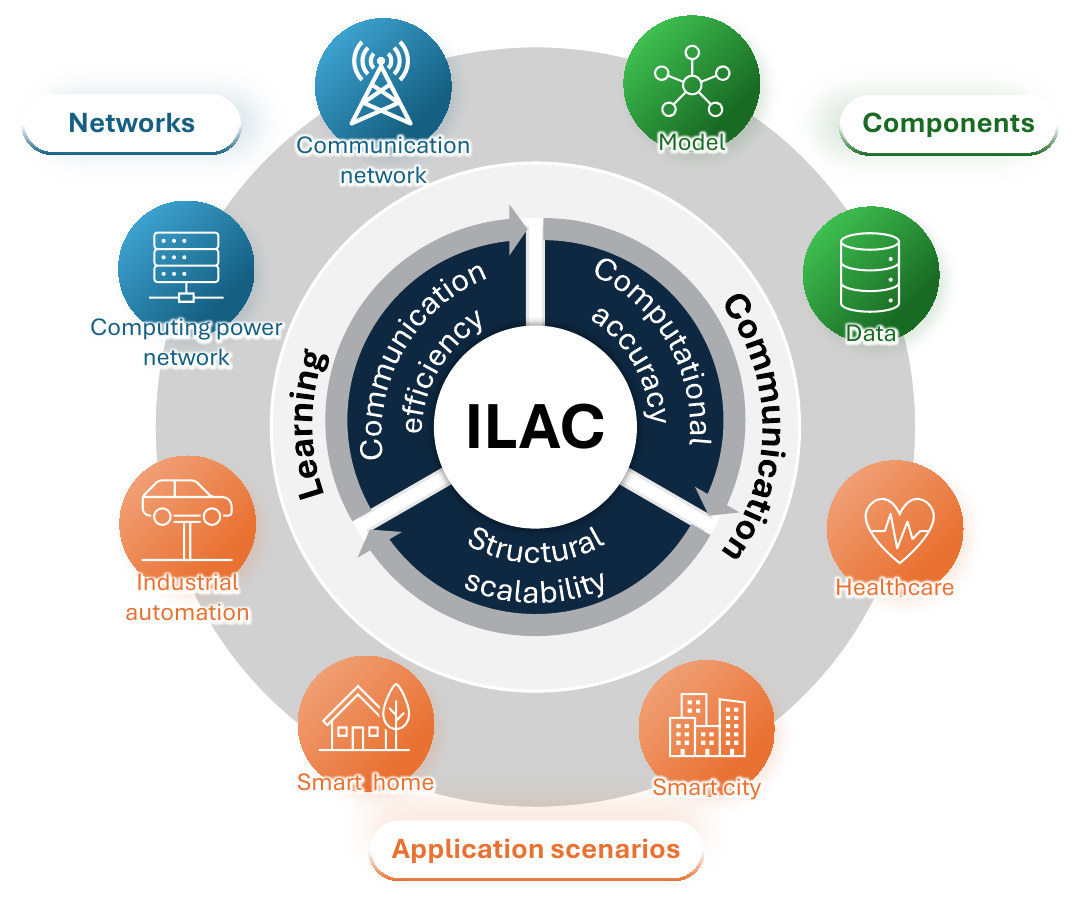}
\caption{Illustration of the unified ILAC framework.} 
\label{ILAC system}
\end{figure}

\item \textbf{Computational accuracy}: Computational accuracy in the ILAC framework pertains to its ability to maintain consistently high-quality learning performance by ensuring both precise model training and robust inference, even under varying resource constraints. This attribute is particularly vital for safety-critical applications, including industrial automation and smart healthcare, where accurate and reliable decision-making is imperative. ILAC tackles this challenge by meticulously balancing learning accuracy against system efficiency, ensuring that computational limitations do not compromise task performance. By integrating adaptive learning mechanisms and efficiently managing model updates, ILAC enhances inference precision while optimizing computational resource utilization.

\item \textbf{Structural scalability}: Structural scalability refers to the ILAC framework’s capacity to seamlessly accommodate expanding network sizes, growing device densities, and evolving task complexities without performance degradation. This is achieved by ensuring smooth device integration, supporting heterogeneous computing capabilities, and enabling dynamic resource allocation in large-scale 6G environments. Such flexible adaptability ensures that ILAC systems can efficiently scale to support the demands of high-density and emerging real-time applications, without requiring extensive architectural modifications or restructuring.

\end{itemize}

On the basis of these principles, the ILAC framework comprises  two interdependent functional networks: the communication network and the computing power network, both centered on data and models. 
Specifically, the communication network ensures low-latency and high-reliability data transmission by dynamically adapting to channel conditions, thereby facilitating efficient data distribution. Meanwhile, the computing network provides distributed resource management, efficiently facilitating large-scale model training and lightweight deployment to optimize the learning process. The synergistic integration of these two networks enables seamless data flow and adaptive model deployment, ultimately enhancing both communication efficiency and learning performance within the ILAC framework.

Given the central importance of models and data, ILAC employs two fundamental architectural strategies that define how these elements are utilized across the system: model distribution and data distribution. Both strategies represent distinct yet complementary paradigms for coordinating learning and communication in distributed environments.

\subsubsection{Model Distribution}

Model distribution involves partitioning AI models and distributing their components across multiple network nodes, thereby enabling collaborative processing where each node contributes toward completing the overall task. In the ILAC framework, model distribution optimizes computational resource allocation, empowering the system to efficiently manage large-scale operations while maintaining scalability. Key approaches such as SL, AL, and intelligent transceiver deployment facilitate the distributed execution of learning and communication tasks without requiring centralized data collection and computation. Specifically, SL alleviates computational burdens by dividing the training process, AL enhances prediction accuracy through ensemble-based model aggregation, and intelligent transceiver design leverages ML-driven signal processing to dynamically optimize communication parameters.

\subsubsection{Data Distribution}
Data distribution in ILAC enables decentralized learning by allocating data processing tasks across multiple devices or nodes, thereby significantly enhancing communication efficiency. Instead of transmitting raw data, techniques such as FL, split federated learning (SFL), and MARL facilitate distributed processing, which supports collaborative learning without the need for centralized aggregation. 
For communication-related tasks, DL models are trained locally at different network nodes to perform intelligent functions such as CSI prediction and resource allocation, reducing the need for frequent data transmission and enhancing adaptability to dynamic network conditions.

Collectively, model distribution and data distribution characterize the decentralized nature of ILAC systems from complementary perspectives. Their integration within the ILAC framework enables a seamless convergence of communication and learning functionalities, effectively addressing the dual demands of scalability, efficiency, and intelligence in future wireless networks.


\subsection{Optimization of ILAC Design}
In practice, to fully unlock the potential of ILAC, a systematic methodology for allocating the limited system resources  for communication and learning is needed. To shed light on this, we formulate the ILAC design as a universal optimization problem, jointly considering computation resources, learning algorithms, and transmission design. The objective is to enhance the dual functionalities of learning and communication while rigorously satisfying resource constraints. Thus, we formulate the generalized optimization problem for ILAC design as follows:
\begin{subequations}\label{eq.gop}
    \begin{align}
		\max_{\mathcal{C},\mathcal{L},\mathcal{R}}\quad & \lambda_1 f\left(\mathcal{C},\mathcal{R}\right) + \lambda_2 g\left(\mathcal{L},\mathcal{R}\right)\tag{\theequation}\\
		\text{s.t.}\hspace{1.2em}
        & f\left(\mathcal{C},\mathcal{R}\right) \geq T_\mathrm{C},\\
        & g\left(\mathcal{L},\mathcal{R}\right) \geq T_\mathrm{L},\\
        & h\left(\mathcal{R}\right) \leq T_\mathrm{R},
    \end{align}
\end{subequations}
where $\mathcal{C}$ is the set of variables related to communication, $\mathcal{L}$ is the set of variables related to learning, and $\mathcal{R}$ is the set of variables related to resource allocation. Here, the functions $f(\cdot)$, $g(\cdot)$, and $h(\cdot)$ respectively represent the communication metric, learning metric, and resource consumption metric, while $\lambda_1$ and $\lambda_2$ are the weights for the communication and learning metrics, respectively. Moreover, the constraints in optimization problem \eqref{eq.gop} ensure that the QoS requirements for both communication and learning are satisfied. Specifically, $T_\mathrm{C}$ and $T_\mathrm{L}$ denote the minimum QoS requirements for the communication and learning metrics, respectively, and $T_\mathrm{R}$ represents the total physical resource budget, including important factors such as bandwidth, transmit power, computation power, storage, and energy consumption. Indeed, the considered generalized ILAC optimization problem \eqref{eq.gop} aims at balancing the non-trivial trade-offs between communication efficiency, learning effectiveness, and resource usage, thereby enabling the development of adaptive and efficient communication and learning systems for next-generation networks \cite{yao2025energy,mlsp,e26050394}.

Specifically, the ILAC optimization problem aims to jointly optimize the communication and learning processes while explicitly considering resource constraints. The paradigm naturally leads to a multi-objective framework \cite{9119454,9851837,7015632} to strike an efficient balance between the two metrics. In the literature, there are various methods to address conflicting goals. One efficient approach is to adopt the weighted-sum method \cite{6077796,10702481,7744443}. Specifically, the objective is to maximize a weighted sum of the communication and learning metrics, each capturing a distinct aspect of ILAC system performance. Unlike traditional communication networks, ILAC's energy consumption and latency encompass not only communication but also local learning processes. 
Therefore, modeling an ILAC system requires considering the interdependent relationships between communication/computation (learning) performance and physical resource allocation. To accommodate different application scenarios, the weighting parameters $\lambda_1$ and $\lambda_2$ in the objective function given by \eqref{eq.gop} enable dynamic adjustment of the relative importance of the communication and learning tasks, thereby adapting system performance to varying operational scenarios.

\begin{figure}[t]
    \centering
    \includegraphics[width=.8\linewidth]{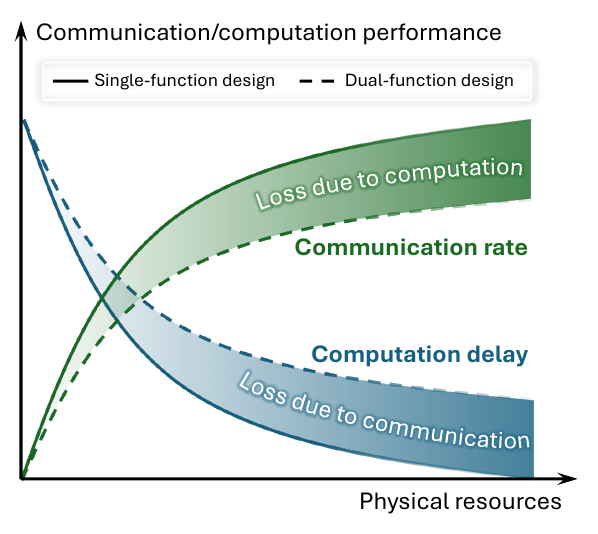}
    \caption{Illustration of the relationship between communication/computation performance and physical resources.}
    \label{relationship between communication/computation performance and physical resources}
\end{figure}

Fig.~\ref{relationship between communication/computation performance and physical resources} illustrates the interplay between the communication and computation performances as a function of the physical resource allocation, leveraging communication rate and computation delay as representative metrics. As the total amount of physical resources continuously increases, communication rates improve, and computation delay decreases, reflecting enhanced system capabilities. In single-function systems optimized exclusively for either communication or computation, all available resources are dedicated to one task, allowing it to approach its theoretical upper performance limit. However, in dual-function systems such as ILAC, resources are shared between learning and communication, introducing inherent competition that prevents either function from reaching its standalone optimum. This fundamental trade-off becomes more pronounced as the total amount of resources increase. Consequently, at high resource levels, the performance gap between ILAC and single-function-optimized systems widens. Notably, this gap is not attribute to inefficiencies in the joint design, but rather arises because resource sharing inherently imposes upper bounds on individual performance objectives.
While resource sharing between learning and communication introduces inherent trade-offs, ILAC achieves mutual reinforcement through adaptive mechanisms. For instance, FL reduces raw data transmission while improving model generalization via collaborative training; SL optimizes model partitioning to balance edge-device computation and server communication; and dynamic resource allocation leverages real-time channel states to enhance both inference accuracy and spectral efficiency. These synergies highlight scenarios where joint optimization enables learning to refine communication strategies and communication efficiency to accelerate learning convergence.


\begin{figure}[t]
    \centering
    \includegraphics[width=.8\linewidth]{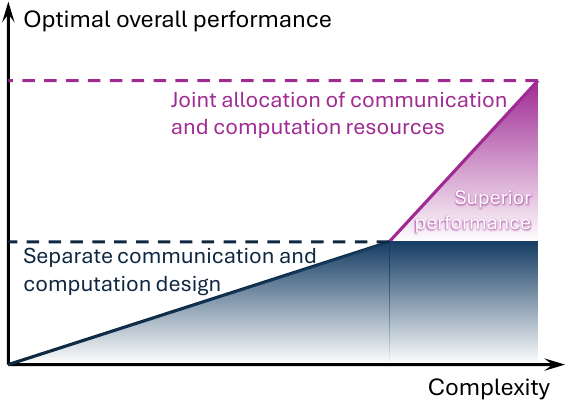}
    \caption{Overall performance versus complexity with different resource allocation schemes.}
    \label{Overall performance versus complexity}
\end{figure}

Fig.~\ref{Overall performance versus complexity} provides additional insights by contrasting single-function and ILAC systems across varying complexity levels. In single-function systems, performance improvements exhibit a steady increase as complexity grows, benefiting from their lower design overhead. In contrast, ILAC systems leveraging joint resource optimization achieve more rapid performance gains with increasing complexity, ultimately surpassing single-function approaches in overall effectiveness. However, merely increasing physical resources does not always guarantee sustained performance improvements in ILAC. The advantages of joint optimization are accompanied by additional system complexity, which introduces additional latency and computational overhead that can offset some performance gains. Therefore, unlocking the full potential of ILAC requires generally not only greater resource availability but also well-designed optimization strategies that balance performance improvements with manageable system complexity. This motivates further research.

\section{Model Distribution in ILAC}\label{Sec:mdai}
Leveraging the enhanced communication and computational capacities of distributed nodes, distributed modeling structures are employed to improve both communication efficiency and computational accuracy. In this section, several model distribution-assisted designs for ILAC are discussed, including SL, AL, and intelligent transceivers. 
\subsection{Learning-oriented Model Distribution Design}
\subsubsection{Split Learning}

\begin{figure*}[t]
    \centering
    \includegraphics[width=.7\linewidth]{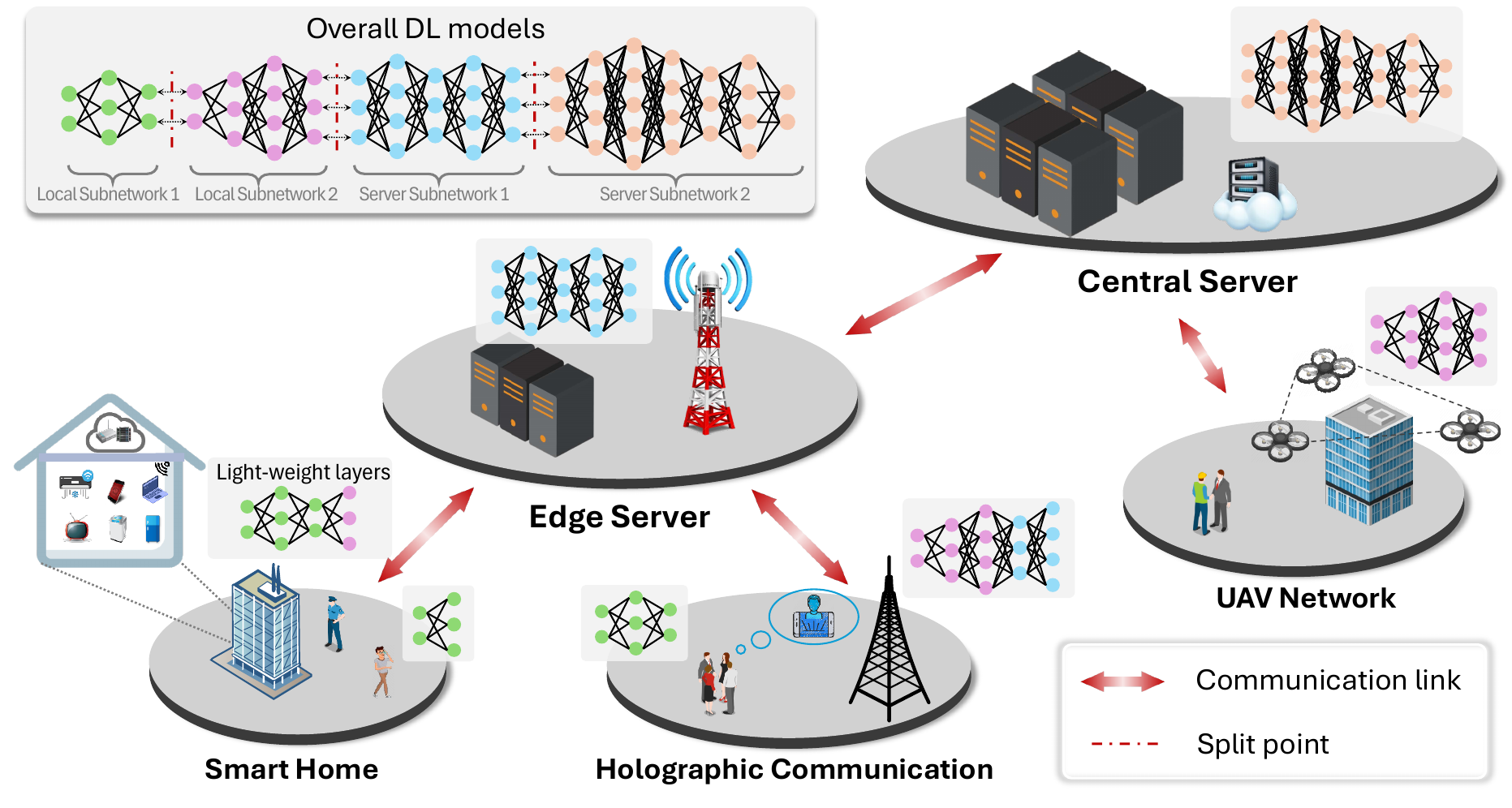}
    \caption{Architecture of SL for heterogeneous wireless networks.} 
    \label{fig:splitlearning}
\end{figure*}


As illustrated in Fig.~\ref{fig:splitlearning}, the inherently distributed nature of infrastructures in wireless networks enables SL, where a deep neural network (DNN) is partitioned across multiple entities, such as client devices, edge servers, and a central server. In this configuration, the computationally intensive layers are offloaded to the central server, which also handles aggregation. In practice, the edge server generally processes layers of moderate complexity due to its intrinsically higher signal processing capability, while the lightweight layers and raw data remain on client devices. Furthermore, wireless communication occurs among these entities, facilitating efficient model training and inference.
One of the key advantages of the SL architecture is the significant reduction in computational overhead for resource-constrained devices. Moreover, by transmitting only intermediate feature representations instead of raw data, the application of SL significantly reduces the regained communication overhead. This leads to more efficient use of bandwidth and lower transmission latency, which are particularly beneficial in bandwidth-limited and high-latency wireless environments. 

However, despite its advantages, SL also introduces several challenges. A fundamental issue is determining the optimal split point of the DL model~\cite{lin2024split}. Indeed, the choice of split point significantly affects both computation and communication overheads~\cite{10504775}. If the model is split too early, the client-side computation may be minimal, but the frequent interactive communications with the server will inevitably incur high communication costs. Conversely, if the split is made too late, the computational burden on the client device increases, defeating the purpose of offloading computations to the server. Furthermore, communication overhead can still be substantial in large-scale systems, especially when frequent exchanges of intermediate feature representations between the client and the server are required. Moreover, the split model also introduces a dependency on the quality and stability of the communication link, which can be problematic in unreliable wireless network conditions. Additionally, due to the openness of the wireless medium, privacy concerns related to potential eavesdropping may arise if sensitive intermediate representations are not properly protected~\cite{10138067}.

To address these challenges, the optimization process for determining the split point takes into account the following key factors:
\begin{itemize}
    \item \textbf{Computation capabilities:} Given the heterogeneous nature of wireless networks, entities such as devices and servers typically exhibit heterogeneous computational capabilities, as illustrated by the distributed models with different scales in Fig.~\ref{fig:splitlearning}. Thus, split-point optimization should ensure an appropriate allocation of computation tasks across entities, maximizing resource utilization and minimizing training latency.
    \item \textbf{Communication overhead:} The transmission of intermediate features between devices and servers introduces certain latency and increases bandwidth consumption. Therefore, efficient communication is critical to prevent excessive data communication overhead from offsetting the benefits of distributed computing.
    \item \textbf{Privacy preservation:} To preserve privacy, the splitting strategy should ensure that only necessary intermediate features are shared. This approach minimizes the risk of exposing private information throughout  distributed model training. 
\end{itemize}

To optimize the individual split point selection for DNNs across heterogeneous wireless devices, efficient model training algorithms have been employed, including integrating FL, parallel computation mechanisms, and environment-aware adaptive splitting. These techniques are detailed as follows. 

Integrating SL with FL capitalizes on the complementary strengths of both approaches~\cite{shiranthika2024optimizingsplitpointserrorresilient,xu2023accelerating,thapa2022splitfed,liu2022wireless,wu2024joint,han2023federated}. Specifically, a split FL framework was applied in \cite{xu2023accelerating} for heterogeneous devices, where split-point selection and bandwidth allocation were jointly optimized to minimize system latency. Additionally, to concurrently achieve both personalization and generalization across different applications, a joint edge-AI training and inference strategy was proposed in \cite{han2023federated}. This strategy not only ensures efficient inference, but also dynamically adapts to network resource availability during the testing phase.

Building upon SL and  its integration with FL, recent research has focused on enhancing the efficiency of SL through parallel computing and innovative communication strategies. 
In particular, the authors of \cite{lin2024efficient} introduced a parallel SL (PSL) framework that seamlessly integrates edge computing with parallel processing techniques to accelerate model training. In this framework, per-round training latency is minimized through the joint optimization of subchannel allocation, power control, and cut layer selection. Additionally, in a wireless multiple-input multiple-output (MIMO) network employing SL, over-the-air computation (OAC), an efficient parallel computation method, was utilized to reduce communication signaling overhead and improve computation efficiency \cite{yang2023over}.  
Such advances in parallel computing offer significant improvements in handling communication and computation demands in SL-based systems.

Compared to the fixed split-point methods mentioned above, adaptive split-point selection has emerged as a promising approach, as it takes into account device heterogeneity and variations in channel conditions.
In particular, by dynamically adjusting model partitions based on the computational capabilities and network conditions of different devices, resource utilization and system efficiency were jointly optimized~\cite{li2024adaptive}. This dynamic adaptation of the split point helps strike an effective balance between computational load and communication overhead. Furthermore, adaptive splitting allows for more efficient trade-offs between computation and communication, enabling the model to be tailored precisely to both client-side tasks and server-side generalization. Consequently, this flexibility ensures that the model not only adapts to variations in device capabilities but also to the number of participating devices, improving the scalability and robustness of ILAC systems. 

Despite these advancements, current SL implementations still encounter several technical challenges. One major issue is the inefficiency in determining the optimal split point, which often leads to imbalanced computational loads, high communication overhead, and increased latency. Additionally, incorporating privacy-preserving techniques remains a complex task, as ensuring robust and scalable privacy protection continues to pose formidable challenges. To address these difficulties, recent works have explored leveraging large AI models, which enable more flexible and dynamic neural network partitioning due to their extensive parameter sets and robust learning capabilities. In fact, this high flexibility allows adaptive adjustments during model partitioning, facilitating effective split-point adaptation  
based on both client-side and real-time network conditions~\cite{ chen2024adaptivelayersplittingwireless, zhang2025splitllmhierarchicalsplitlearning}. Such adaptability significantly improves the optimization of split-point selection by distributing the computational load more effectively across devices and servers. Additionally, the exploitation of large AI models can bolster the accuracy and efficiency of privacy mechanisms. Specifically, by processing vast amounts of data and extracting high-level semantic features, these models can implement advanced privacy-preserving algorithms, such as differential privacy and homomorphic encryption, ensuring data security without compromising model performance. 

\subsubsection{Aggregation Learning}
Building upon the advantages of distributed processing, AL combines predictions from multiple distributed models to improve accuracy, robustness, and generalization. This inherent distribution of terminals in wireless networks complements the distributed nature of AL, thereby highlighting its significant potential for implementation in practical wireless communication systems. 
However, several critical challenges must be carefully addressed, including the dynamic variability of wireless communication environments and communication security provisioning. 

Firstly, the mobility of communication nodes introduces continuous random variations in network connectivity. For example, sensor nodes in aquatic wireless sensor networks (WSNs) are in continuous motion due to water currents or shifts in wind direction. Specifically, in underwater WSNs, nodes drift from their original positions with ocean currents, rendering timely adaptation a significant challenge. Moreover, environmental factors such as fluctuations in temperature, humidity, and pressure can adversely impact the propagation of wireless signals, thus degrading the quality of data transmission. Indeed, these dynamic changes require that AL models continuously ensure effective data collection and transmission. Similarly, changes in network topology directly influence data routing and distribution. For instance, the frequent addition or removal of nodes in WSNs leads to constant variations in the network's topology, placing higher demands on the stability and adaptability of the AL models. Furthermore, concept drift \cite{jain2024instance} is another critical challenge that AL must address in dynamic environments. In practice, data distribution evolves over time within the network, rendering previously learned patterns inapplicable. Consequently, AL models must be robust enough to continuously update and adapt to these changes.

Moreover, security and privacy are paramount considerations when implementing AL in practical wireless communication systems. Various threats, such as adversarial attacks, eavesdropping, and data manipulation, can compromise the integrity and confidentiality of aggregated data. Specifically, adversarial attacks involve malicious attempts to manipulate the aggregated learning process by injecting false data, tampering with transmission signals, or compromising individual nodes to alter the aggregation outcomes. Furthermore, eavesdropping refers to the unauthorized interception of wireless transmissions, where eavesdroppers may intercept and analyze the exchanged data or signals to infer sensitive information. Additionally, since AL relies on the collection and processing of data from multiple sources, malicious actors may manipulate or corrupt the data during transmission or storage, resulting in potentially inaccurate aggregation results and thus jeopardizing both data integrity and individual privacy.


\begin{figure*}[t]
    \centering
    \includegraphics[width=.8\linewidth]{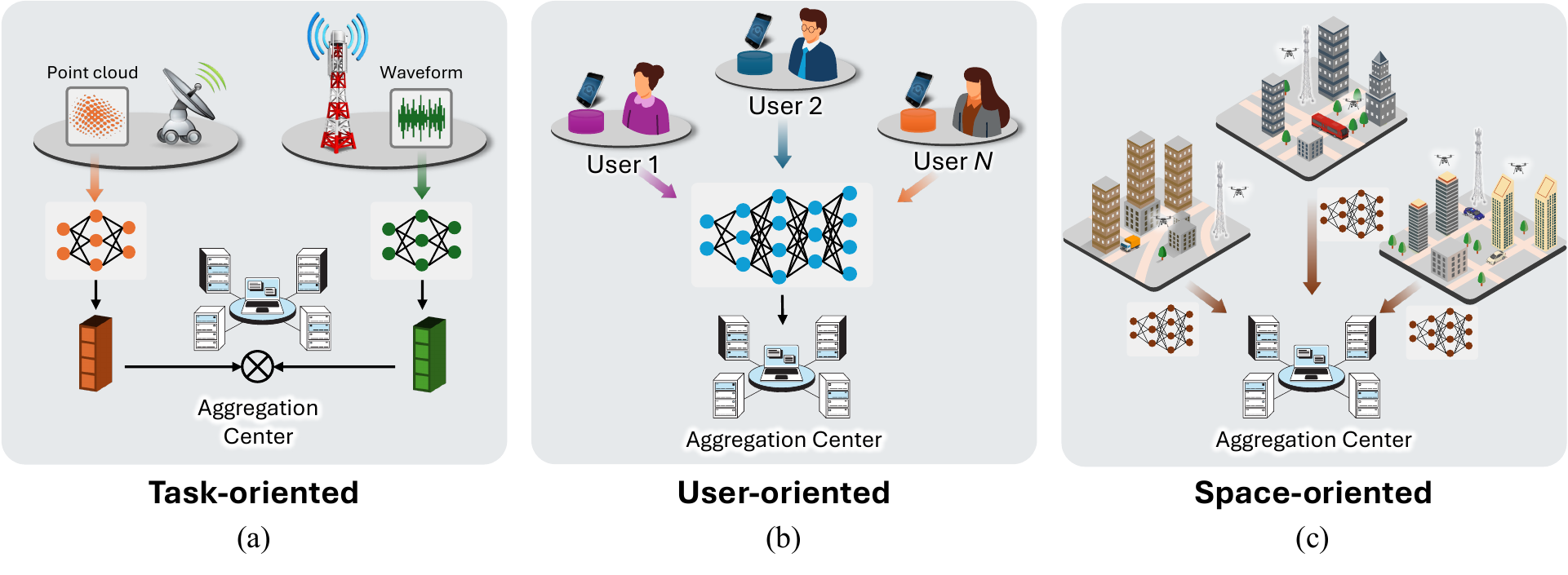}
    \caption{Three main paradigms of aggregation learning in practical wireless communication scenarios.} 
    \label{fig:aggregationlearning}
\end{figure*}

To address these dynamic and security-related challenges inherent in distributed AL systems, various pragmatic aggregation paradigms have been developed, tailored specifically to different application scenarios and objectives. These approaches strategically coordinate learning tasks, facilitate user collaboration, and enhance spatial adaptability to mitigate environmental uncertainties while safeguarding against adversarial threats. Specifically, three main aggregation paradigms have emerged: task-oriented aggregation, user-oriented aggregation, and space-oriented aggregation, as shown in Fig.~\ref{fig:aggregationlearning}.

\begin{itemize}
    \item \textit{The task-oriented aggregation paradigm} illustrated in Fig.~\ref{fig:aggregationlearning}(a) combines different learning tasks by exploiting their inherent complementarity to enhance overall inference and prediction capabilities. For instance, a novel multi-cue fusion (MCF) network was proposed in \cite{9427151} for automatic modulation recognition. The MCF network consists of a signal-cue multi-stream (SCMS) module and a visual cue discrimination (VCD) module. Specifically, the SCMS module is designed for capturing spatial-temporal correlations from two distinct signal cues, aiming to explore differences and harness the complementary information from multiple data forms. In parallel, the VCD module converts raw I/Q data into constellation diagrams that serve as visual cues to extract critical structural information. Subsequently, a score fusion method is exploited to combine the outputs from both tasks, resulting in more reliable predictions. However, the parallel design of multiple modules leads to a large number of model parameters, which increases the demand for computational resources. Moreover, the data pre-processing steps are typically cumbersome and resource-intensive, which may limit the real-time applicability of this approach. 
    \item \textit{The user-oriented aggregation paradigm} consolidates data or models from different users, leveraging the diversity among them to bolster system robustness. This diversity can be particularly useful in resisting adversarial attacks, data manipulation, and other malicious methods, especially when certain users' communication or data collection processes are compromised. For instance, in \cite{9127944}, a robust cooperative spectrum sensing framework based on ensemble machine learning (EML) was proposed for full-duplex cognitive radio networks (FD-CRNs), demonstrating high accuracy and strong resilience against malicious attacks and interference. However, the integration of multiple models inevitably leads to a significant increase in computational complexity. 
    \item \textit{The space-oriented aggregation paradigm} shown in Fig.~\ref{fig:aggregationlearning}(c) focuses on learning differences across data collected from distinct regions or environments. This approach enables rapid adaptation to dynamic and time-varying communication environments, thereby enhancing the adaptability and predictive accuracy of aggregation models in new scenarios. For example, the authors in \cite{9845684} proposed a novel deep ensemble learning-based, mobile-network-assisted UAV monitoring and tracking system for detecting spoofing attacks on cellular-connected UAVs. The proposed method deployed deep ensemble learning at edge cloud servers and utilized multi-layer perceptron (MLP) neural networks to analyze path loss statistical features of base stations (BSs) at different locations to make a final decision. However, due to the significantly longer training time of the MLP model compared with simpler ML models, this method may introduce increased latency when dynamically updating the model in response to environmental variations.
    
\end{itemize}

In summary, \textit{task-oriented aggregation} enhances adaptability by combining different learning tasks, enabling the system to maintain performance despite varying network conditions while simultaneously creating a more complex attack surface by combining multiple tasks, making it harder for adversaries to manipulate the system; \textit{user-oriented aggregation}, on the other hand, leverages diversity among users to improve robustness against fluctuating connectivity and strengthens security through diversity, since compromising individual users becomes less effective against the aggregated system; \textit{space-oriented aggregation} focuses on learning differences across regions, allowing rapid adaptation to dynamic communication environments and reducing vulnerability by focusing on localized data patterns rather than centralized data.
\subsection{Communication-oriented Model Distribution Design}
Recently, DL algorithms have demonstrated significant effectiveness in optimizing the higher layers of the communication stack, thereby advancing the development of intelligent transceivers. Specifically, intelligent transceivers adaptively adjust key communication parameters, such as channel coding, modulation, and beamforming, to optimize transmission performance through real-time CSI estimation and other environmental conditions. 

Despite these advancements in DL-based intelligent transceiver design, two main challenges remain: Accurate channel modeling and robustness under realistic channel conditions. In practice, actual communication channels comprise various nonlinear components, such as digital or analog pre-distortion and conversion, as well as non-differentiable stages like up-sampling and down-sampling. As a result, the deep neural layers in intelligent transceivers are often trained using constructed or simplified channel models rather than the actual channel conditions, which can lead to performance degradation during inference. Moreover, since all hidden layers and intermediate layers are trained based on the posterior probability of the input signals, any distortions affecting the initial layers can propagate through the network. In practical intelligent transceivers, the first deep neural layer at the receiver is particularly vulnerable, as its inputs are directly influenced by real-time continuous channel distortions. This vulnerability can cascade through subsequent layers, potentially leading to overall reasoning failure. As such, addressing these issues is critical for enhancing the robustness of DL models in intelligent transceiver applications.

Due to the inherent characteristics of DL, there is a noticeable lack of interpretability among neural layers \cite{zhang2021survey}. Hence, it is impossible to precisely identify which neurons and connections between layers most significantly influence the overall learning accuracy. Indeed, deep neural network-based classifiers often rely on certain critical paths for final decision-making. In particular, if these critical paths remain intact, accurate classifications can be achieved. However, if the critical paths are disrupted, it can lead to erroneous outcomes \cite{zhang2022deep}. Moreover, deep neural networks are highly susceptible to noise, a potentially critical limitation for their application in intelligent transceivers. This underscores the need for enhancing the robustness of intelligent transceivers against transmission noise. 


\begin{figure*}[t]
    \centering
    \includegraphics[width=.85\linewidth]{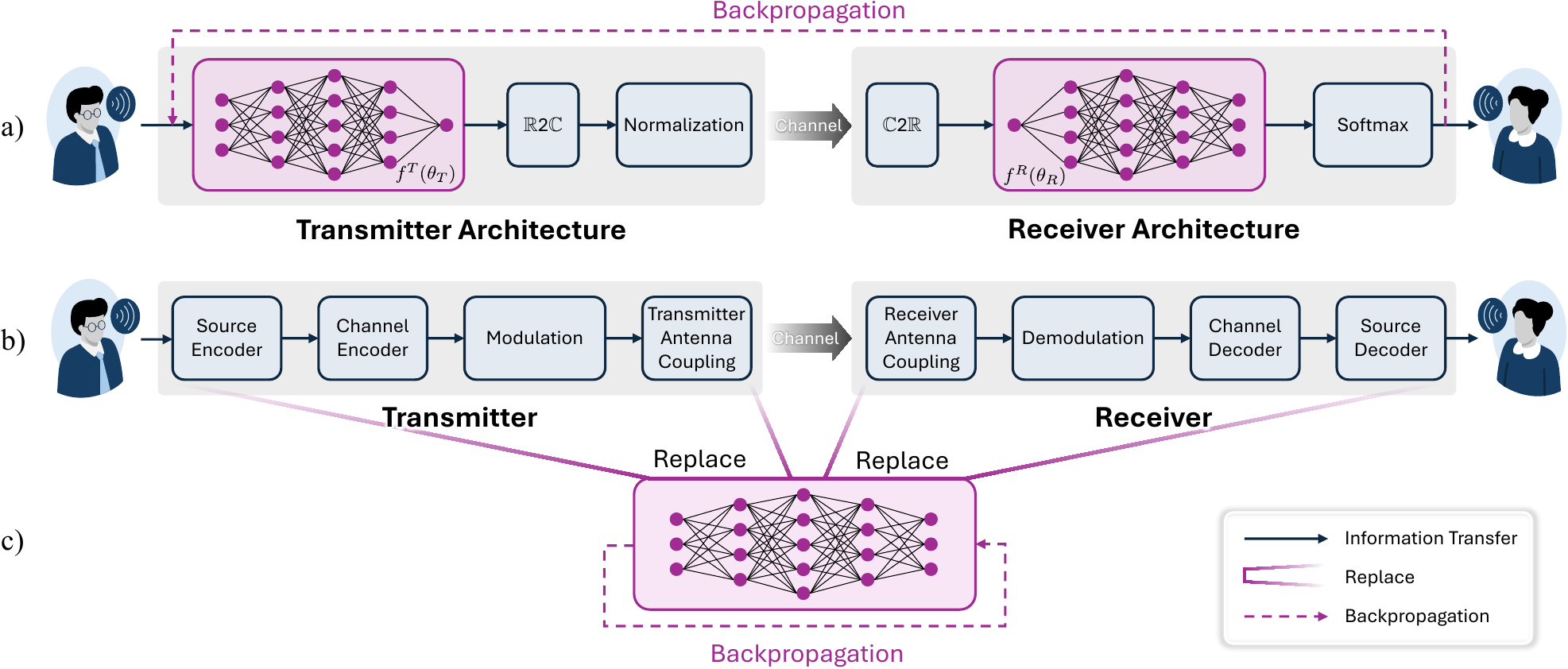}
    \caption{Structures of a conventional wireless communication system and intelligent transceiver: (a) Symmetrical intelligent transceiver; (b) Conventional wireless communication system; (c) Asymmetrical intelligent transceiver.} 
    \label{fig:semcom}
\end{figure*}

The aforementioned challenges in DL-driven transceiver robustness and model interpretability have motivated the exploration of structural designs that reconcile learning adaptability with the dynamic channel conditions. From a unified viewpoint, there are two distinct paradigms for designing wireless intelligent transceivers: the symmetrical paradigm \cite{8985539, 8645416} and the asymmetrical paradigm \cite{8768319, 9775110, 9322310, 8052521,gu2025task}. As shown in Fig.~\ref{fig:semcom}(a), the symmetrical paradigm employs deep neural network models symmetrically designed at the transmitter and the receiver. In this approach, the neural layers at both ends are jointly optimized, adapting simultaneously to the source and the target in a specific channel environment. In contrast, the asymmetrical paradigm focuses on modeling specific intelligent transceiver modules as deep neural networks to improve performance, e.g., by optimizing beamforming at the transmitter or channel detection and estimation at the receiver. In the following, we further analyze the technical foundations and implementation strategies of these two paradigms, demonstrating how their distinct design philosophies address the fundamental trade-offs between channel-agnostic generalization and task-specific optimization in intelligent transceiver systems.

\subsubsection{Symmetrical Intelligent Transceiver} Due to the inherent random variations and non-differentiable stages of wireless channels, it is often challenging to model real-world systems adopting differentiable DL models. In fact, even state-of-the-art channel models cannot fully capture all possible scenarios of real-world radio propagation \cite{wang2023doppler,wang2025flexible,zhang2025closer}. Consequently, the generalization capability of end-to-end learning in traditional communication systems is inherently limited. Considering these challenges in obtaining CSI, symmetrical paradigms have been devised to achieve effective end-to-end learning in scenarios where the channel model is unknown. For example, a novel learning algorithm was proposed in \cite{8645416} to achieve end-to-end learning in communication systems leveraging neural network-based autoencoders. This method demonstrates that autoencoders can be trained solely from raw observations without relying on prior knowledge of the underlying channel model. Additionally, a conditional generative adversarial network (GAN) was utilized in \cite{8985539} to model channel effects and bridge the transmitter and receiver DNNs, enabling gradient back-propagation from the receiver DNN to the transmitter DNN. This approach paved the way for automatic learning of channel effects in intelligent transceivers without requiring knowledge of the specific channel transfer function. However, the current symmetrical intelligent transceiver paradigm continues to encounter several common limitations, including high training complexity, considerable resource consumption, as well as dependency on idealized feedback mechanisms. 

\subsubsection{Asymmetrical Intelligent Transceiver} The asymmetrical paradigm illustrated in Fig.~\ref{fig:semcom}(c) is adaptively designed to respond to random variations inherent in practical wireless channels. By learning and optimizing essential communication parameters of the transceiver, this paradigm is powerful enough to effectively cope with complex channel environments, enhancing both communication quality and resource utilization. For instance, on the transmitter side, a novel framework of low-cost link adaptation for spatial modulation multiple-input multiple-output (SM-MIMO) systems was proposed in \cite{8768319} based upon machine learning. In this framework, the problems of transmit antenna selection (TAS) and power allocation (PA) in SM-MIMO were transformed into data-driven prediction tasks, departing from traditional model-based optimization methods. The approaches proposed in \cite{8768319} were further extended to other adaptive index modulation (IM) schemes, demonstrating that DNN-based designs can consistently outperform various conventional optimization-driven approaches while achieving excellent performance with significantly lower complexity. Additionally, the authors of\cite{9322310} proposed a convolutional massive beamforming neural network (CMBNN) exploiting both supervised and unsupervised learning schemes, with a specific design for the network's structure and input/output. On the receiver side, a DL-based model for channel estimation and signal detection was proposed in \cite{8052521} for orthogonal frequency-division multiplexing (OFDM) systems. The proposed DL-based approach implicitly estimated CSI and directly recovered the transmitted symbols, effectively addressing channel distortion and exhibiting greater robustness, particularly in practical scenarios with nonlinear clipping noise. In parallel, distributed machine learning (DML) techniques were leveraged to enable reliable downlink channel estimation \cite{9775110}, facilitating the  extraction of different channel features for different channel scenarios. However, the current asymmetric intelligent transceiver paradigm faces persistent limitations, such as data dependency, generalization bottlenecks, and the fragility associated with implicit CSI estimation methods.

\section{Data Distribution in ILAC}\label{Sec:ddai}
This section explores data distribution methodologies and their applications in ILAC focusing on both learning-oriented and communication-oriented perspectives. In learning-oriented designs, distributed learning paradigms, such as FL and MARL, are employed to facilitate communication. Conversely, communication-oriented designs leverage data distribution techniques, including federated CSI compression and federated channel prediction, to optimize communication performance.

\subsection{Learning-oriented Data Distribution Design}
\subsubsection {Federated Learning}
FL enables multiple clients to collaboratively train a global model by leveraging their local data, while maintaining raw data privacy on-device \cite{ni2022star,ni2023semi,yao2025byzantine}. This setup not only safeguards data privacy but also supports parallel training, establishing an efficient framework for distributed model development. However, the real-world deployments of FL present two critical challenges.

\begin{figure}[t]
    \centering
    \includegraphics[width=1\linewidth]{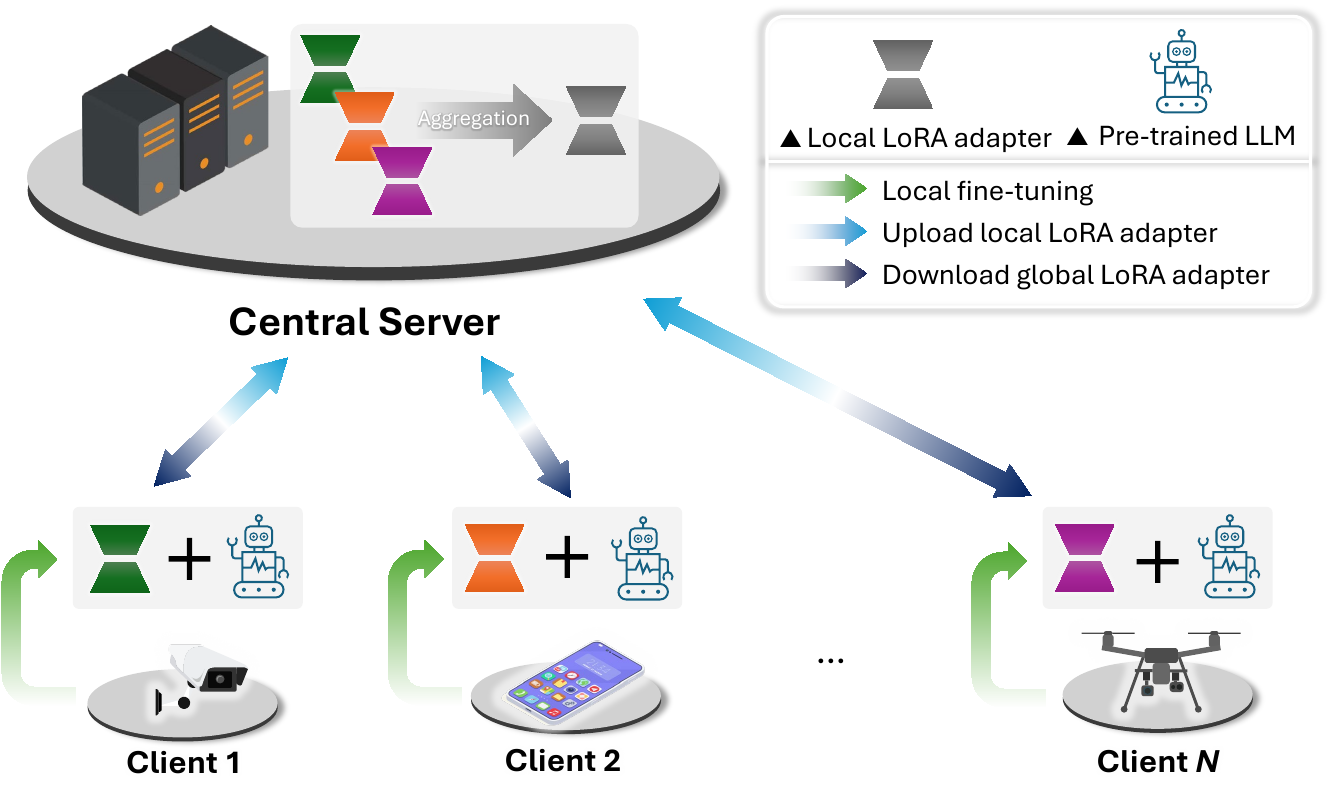}
    \caption{Workflow of low-rank adaptation fine-tuning for large models based on federated learning.}
    \label{fig:FL}
\end{figure}



First, as model sizes increase, transmitting complete model parameters and executing local training can become computationally demanding and communication-intensive, posing significant challenges for edge devices. To mitigate these bottlenecks, parameter-efficient training techniques, such as low-rank adaptation (LoRA), have emerged. Specifically, as illustrated in Fig.~\ref{fig:FL}, LoRA addresses these challenges by freezing pre-trained parameters and applying low-rank decomposition to the learnable updates, thereby significantly reducing the volume of parameters that need to be transmitted~\cite{jiang2023low}. Furthermore, to improve communication efficiency,  LoRA can be combined with mixed-precision quantization strategies, which dynamically adjust parameter bit widths based on their sensitivity. These mixed-precision quantization strategies optimize communication cost and accelerate learning convergence ~\cite{yi2023fedlora}. The combined strategy dynamically allocates quantization bit numbers based on parameter sensitivity and employs a nonlinear quantization table to maximize information entropy.

\begin{figure}[t]
\centering
\includegraphics[width=1\linewidth]{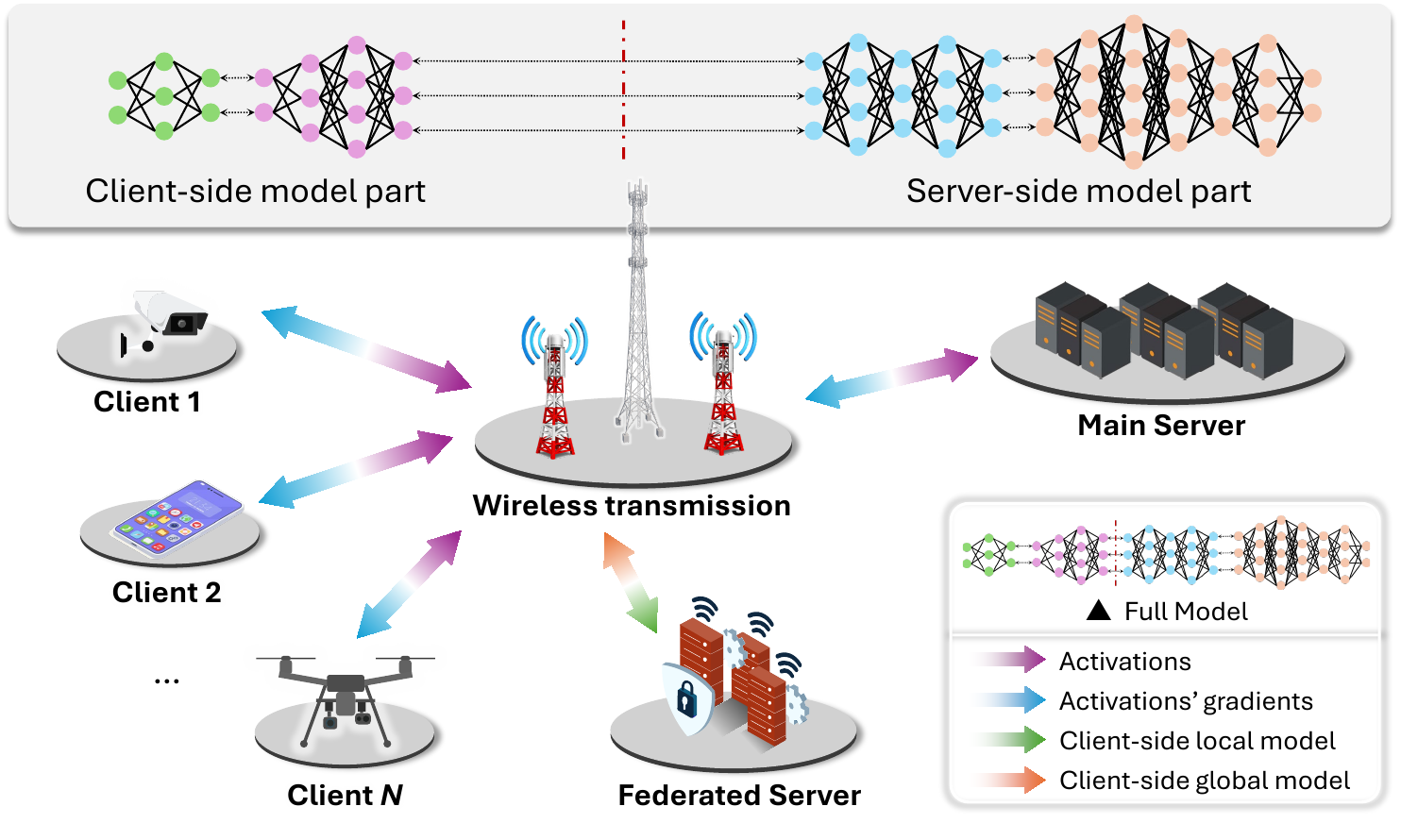}
\caption{Architecture and workflow for split federated learning over wireless networks\protect\footnotemark.}
\label{fig:SFL}
\end{figure}

Second, numerous edge devices process limited computational capacity to perform complete model training. To address this limitation, as shown in Fig.~\ref{fig:SFL}, SFL partitions the model between clients and the server, allowing clients to train only shallow layers while relying on a central server for deeper computations \cite{thapa2022splitfed}. Specifically, a federated server coordinates and aggregates models across clients, making SFL partially feasible in resource-limited environments~\cite{lin2024splitlora}. While SFL effectively reduces the local computational burden, it introduces substantial communication signaling overhead due to frequent exchanges of intermediate features and gradients. To counteract this, recent research has focused on the joint optimization of communication and computation resources, employing techniques such as multi-objective scheduling ~\cite{Zhao2024FedsLLMFS} and Lyapunov-based decision-making~\cite{battiloro2022lyapunov} to dynamically balance performance and efficiency.

\footnotetext{It is a \textit{data distributed system}, since different clients only own part of the whole data of the system.}

The emergence of large AI models provides a promising approach for addressing several existing challenges in distributed learning. Their strong generalization abilities mitigate the impact of non-IID client data distributions, thereby reducing their reliance on large and diverse datasets for training. Additionally, parameter-efficient fine-tuning methods, such as LoRA, enable scalable deployment by significantly lowering the communication overhead. Furthermore, large models facilitate transfer learning across clients and tasks, accelerating convergence and reducing the number of communication rounds required. When paired with architectural decoupling strategies, such as SFL, large models can be flexibly deployed across heterogeneous devices and servers, thus significantly improving the scalability, adaptability, and overall efficiency of ILAC systems.

\subsubsection{Multi-Agent Reinforcement Learning} 
In recent years, MARL has emerged as a prominent research area in AI due to its effectiveness in distributed collaborative decision-making. With the rapid growth of IoT and 5G networks, there is an increasing demand for distributed learning and real-time communication collaboration in complex systems, such as smart grids and the IoV. Consequently, this has led to a deeper integration of MARL into advanced communication system designs. MARL leverages the distributed training data from multiple agents to collaboratively train a global model in a distributed manner \cite{tan1993multi,littman1994markov}.
However, due to the spatially distributed nature of observable data in typical MARL systems, several core challenges arise:
\begin{itemize}
\item \textbf{Communication-computational resource coupling:}  
Agents must strike an effective balance between computationally intensive local training, frequent policy synchronization, and constrained communication resources. For example, in dynamic spectrum access, distributed base stations perform Q-table synchronization for spectrum allocation; however, excessive communication among agents can induce congestion and increase latency.
\item \textbf{Non-independently and identically distributed data:} Data distributions observed locally are influenced by factors such as geographic locations and environmental dynamics, potentially resulting in locally optimal policy updates rather than global optimality. 
\item \textbf{Communication reliability constraints:} Wireless channel fading, interference, and device mobility can cause intermittent communication link disruptions, posing reliability challenges for multi-agent coordination.
\end{itemize}

In response to these challenges, extensive research in learning-oriented design has been initiated. For example, a fully decentralized MARL framework was proposed in \cite{chen2024communication} to implement randomly rounded numerical quantization, transmitting only nonzero components post-quantization to significantly reduce communication overhead while maintaining control performance. Furthermore, in \cite{hua2024communication} a communication-efficient MARL framework was designed for collaborative adaptive cruise control for connected and autonomous vehicles, integrating quantized stochastic gradient descent and binary differential consensus to enhance fleet stability and energy efficiency. Moreover, for fully cooperative MARL scenarios, where agents may receive excessive irrelevant information, the authors of \cite{zhu2024hypercomm} introduced a hypergraph-based communication framework. This model structured multi-agent interactions via hyperedges, enabling more efficient and precise inter-agent communication.

In the ILAC framework, agents must concurrently optimize their local strategies while adapting to the dynamic behaviors of other agents, inevitably escalating computational and communication complexity. By jointly optimizing communication and computing processes, the overall efficiency and performance of the system can be significantly enhanced. Moreover, as large-scale generative AI models are embedded in wireless networks, MARL must address the emerging challenges associated with decentralized storage and synchronization of model parameters. This represents a critical research direction for the development of intelligent 6G networks.

\subsection{Communication-oriented Data Distribution Design}

In practical systems, data distribution design emphasizes localized multi-node data processing to achieve global learning objectives in decentralized architectures. Traditional centralized communication systems often encounter various challenges, such as high transmission latency and increased privacy risks. In contrast, data distribution strategies enhance data privacy and model generalization by leveraging autonomous node training and collaborative learning mechanisms \cite{niknam2020federated}. This subsection further investigates the pivotal role of data distribution in enabling multi-node collaboration within communication networks.

In data distribution designs, each communication node independently trains its local model exploiting heterogeneous local data, such as channel measurements and resource usage records. Knowledge sharing among nodes is achieved through lightweight parameter exchanges, enabling the system to adapt to wireless network variations while maintaining low-latency requirements \cite{chafii2023emergent}. This decentralized approach provides a scalable and efficient solution for distributed learning and communication tasks, enhancing both learning accuracy and communication efficiency.


\subsubsection{CSI Prediction}
Efficient and cost-effective CSI acquisition is essential for enabling intelligent cellular networks with ubiquitous access and heterogeneous device support. In fact, CSI prediction can be formulated as a data-driven DL task. {Conventionally, local CSI prediction methods rely on historical data. For instance, a single base station can leverage a local long short-term memory (LSTM) network to predict future CSI variations based on past channel measurements. This local prediction approach effectively eliminates the communication overhead associated with transmitting data across nodes.} However, offline-trained neural network models often face challenges, such as data silos and limited online adaptability. To overcome these challenges, FL provides a decentralized framework for continuous CSI prediction, offering robustness against inter-cell heterogeneity \cite{sun2023interacting}. Specifically, by transmitting model parameters instead of raw CSI data, FL dramatically reduces communication signaling overhead compared to centralized learning, while achieving comparable performance in CSI estimation and feedback.

\subsubsection{Resource Allocation}
Over the past few decades, advancements in information and communication technologies have primarily focused on enhancing transmission capacity. However, the associated excessive energy consumption has become a significant bottleneck. Additionally, the scarcity of spectrum resources necessitates multiplexing among numerous devices, leading to severe co-channel interference. Consequently, joint optimization of power control and resource allocation has emerged as a critical challenge.

{The traditional approach to resource allocation primarily relies on autonomous optimization at individual nodes. For example, terminal devices can employ Q-learning algorithms to determine transmit power control strategies based on historical local signal-to-noise ratio measurements. By exploiting energy efficiency as a reward function and iteratively exploring power adjustment strategies through a trial-and-error mechanism, the algorithm gradually converges to an effective solution. Although such decentralized methods do not incorporate inter-node collaboration, they provide a solid theoretical foundation for more complex algorithms.} In contrast, centralized DRL methods are often impractical due to the significant delays induced by information exchange, rendering them unsuitable for real-time applications. Furthermore, as the number of devices increases, the computational complexity of centralized schemes escalates dramatically, resulting in substantial computational costs~\cite{ji2023multi}.


\begin{figure}[t]
\centering
\includegraphics[width=1\linewidth]{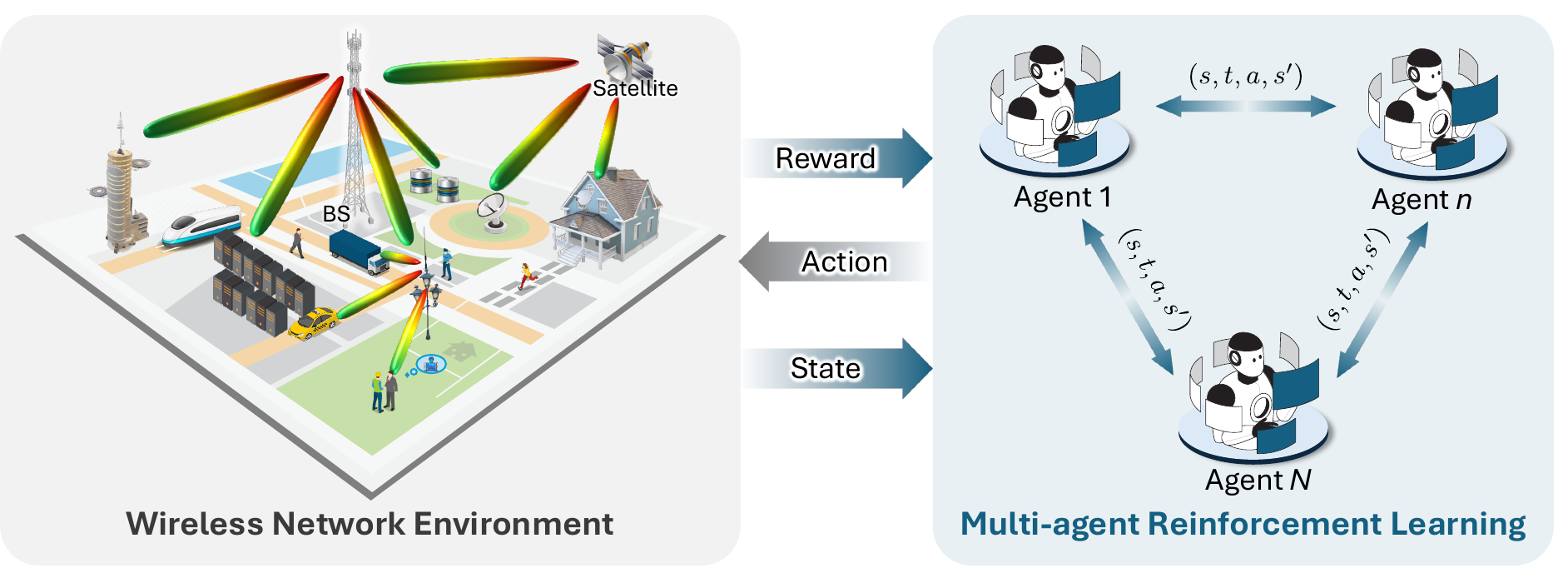}
\caption{Adaptive resource allocation via MARL for wireless communication networks.} 
\label{marl}
\end{figure}

In response to the aforementioned challenges, decentralized MARL algorithms offer significant improvements in communication efficiency. As illustrated in Fig.~\ref{marl}, a MARL system receives state information, e.g., channel conditions, latency metrics, and QoS rewards, from the wireless network environment. Utilizing this information, the system determines optimal actions such as spectrum allocation or channel selection. By enabling collaborative decision-making among agents, MARL effectively mitigates the non-smooth challenges arising from simultaneous agent actions. Furthermore, MARL surpasses purely distributed RL methods by avoiding suboptimal local solutions, thereby achieving superior convergence and enhancing system performance in both low-latency and latency-tolerant applications~\cite{tian2021multiagent,he2023scheduling}.

In addition, reconfigurable intelligent surfaces (RISs) are emerging as a key enabler for 6G wireless networks. An RIS is a planar structure composed of numerous tunable elements capable of dynamically manipulating electromagnetic waves by adjusting their reflection, refraction, and scattering properties. Through real-time control of the phase and amplitude of each element, RISs can reshape wireless propagation environments to enhance signal coverage, suppress interference, and improve energy efficiency. In state-of-the-art RIS-assisted communication systems for 6G, optimized beamforming plays a critical role for improving the performance. Typically, existing beamforming methods employ heuristic algorithms, where each RIS unit dynamically adjusts its phase offset through a greedy strategy based on locally received signal strength \cite{ni2025reconfigurable,shi2022secure,ni2022integrating,shi2024secrecy}. Indeed, centralized learning methods leverage deep neural networks to learn the mapping between the CSI and the optimal RIS phase configuration. However, as the number of connected devices increases, centralized learning approaches introduce significant communication and computational overheads \cite{ni2021federated}. To address this limitation, FL-assisted beamforming methods transmit only model parameters instead of raw data, achieving efficient communication and computation in a distributed manner \cite{shi2025combating}. Moreover, RIS systems with different numbers of reflective elements often experience varying signal propagation conditions and diverse channel characteristics. By leveraging personalized modular architectures and layered designs, FL facilitates effective model training tailored to specific RIS configurations, ensuring high performance while minimizing computational complexity \cite{min2022stratified}.

\section{Computational Complexity and Communication Overhead of ILAC}\label{Sec:ccco}
In this section, we discuss two critical performance metrics in the ILAC framework: computational complexity and communication overhead. Specifically, we examine these performance metrics for popular DL algorithms and identify potential optimization strategies to enhance overall system efficiency. 

\subsection{Computational Complexity}
Computational complexity refers to the computational resources required for both model training and task inference, typically quantified in terms of the number of operations and training duration~\cite{xu2023edge}. For distributed systems operating over heterogeneous wireless networks, managing computational complexity is crucial to determine the feasibility of deploying deep learning models on resource-constrained clients, directly impacting both latency and scalability.

In this context, traditional convex optimization algorithms, which constitute the mathematical backbone of numerous learning formulations, have been extensively studied in terms of their computational cost~\cite{9775110,10024766}. For instance, in~\cite{yang2020energy}, the successive convex approximation (SCA) and Dinkelbach methods were employed for an energy-efficient FL framework, with associated complexities of $\mathcal{O}(N^3 L_\mathrm{SCA})$ and $\mathcal{O}(N\log_2(1/\epsilon))$, respectively. Here, $\mathcal{O}(\cdot)$ refers to Big-O notation, $N$ represents the number of clients, $L_\mathrm{SCA}$ is the total number of iterations, and $\epsilon > 0$ is the required solution accuracy. These algorithms are considered to operate in efficient time, given that their complexity grows polynomially with respect to $N$, while the complexity with with respect to $\epsilon$ grows logarithmically. However, despite their theoretical tractability, the computational burden can still become significant in large-scale deployments or under tight latency and energy constraints. This limitation makes convex optimization methods increasingly insufficient as standalone solutions for practical wireless edge scenarios.

To better understand and address the scalability challenges in ILAC systems, it is essential to examine the computational characteristics of DL models. Unlike convex methods, DL models exhibit diverse and often non-convex behaviors, necessitating a layer- and model-level analysis of their computational loads. Specifically, the computational complexity of CNNs~\cite{10552192,zhang2022deep, 8985539}, which are commonly employed in image processing tasks, is dominated by convolution operations over a feature map of size $m \times m$ with $k$ filters of size $f \times f$, resulting in a per-layer complexity proportional to $\mathcal{O}(km^2 f^2)$. As for sequence tasks, recurrent neural networks (RNNs) have a per-sequence complexity of $\mathcal{O}(t  h^2)$, where $t$ denotes the number of time steps and $h$ is the hidden state size~\cite{zhou2021incorporating}. Regarding graph neural networks (GNNs), increasingly employed for processing graph‑structured data~\cite{xu2023distributed}, message-passing operations across $v$ nodes and $e$ edges incur $\mathcal{O}(vh+eh)$ operations. In the context of large AI models, such as transformers and other foundation models, e.g., GPT and vision transformer, the computational complexity becomes even more pronounced~\cite{zeng2024csigptintegratinggenerativepretrained,edge_llm}. Transformer-based architectures, widely used in natural language processing and increasingly adopted in wireless communication applications, exhibit a computational complexity of $\mathcal{O}(hl^2)$, where $l$ denotes the sequence length.

While the computational complexity of individual models is critical, distributed learning frameworks, such as FL and SL, introduce additional complexity due to task partitioning between clients and servers. Without loss of generality, let $d$ denote the total data size, $s$ the model size, $r$ the communication rate, and $T$ the time required for one forward and backward propagation over the model. The total training time in a typical FL framework can thus be modeled as~\cite{yang2022federated,8664630}
\begin{align}
    \mathcal{T}_\text{FL}=T+2s/r+T_\text{agg},
\end{align}
where $T_\text{agg}$ represents the aggregation time. The factor of $2$ arises because each FL round involves bidirectional transmissions of model parameters between the client and the server. Similarly, in SL, the computational load is distributed between the client and server, yielding a total training time of~\cite{thapa2022splitfed} 
\begin{align}
    \mathcal{T}_\text{SL}=T+2dq/r+2N\beta s/r,
\end{align}
where $q$ is the size of intermediate ``smashed data" transmitted between client and server and $\beta$ represents the fraction of model parameters stored locally.

Comparing the above approaches, traditional convex optimization methods offer analytical convergence guarantees and operate in polynomial time. However, their centralized and iterative nature makes them less practical for large-scale or time-sensitive wireless environments. FL distributes the training workload across multiple clients, with its total time $\mathcal{T}_\text{FL}$ reflecting local computation, communication overhead, and server-side aggregation. This approach is efficient when clients have adequate resources and stable communication links. SL, in contrast, offloads a portion of the computation to the server and transmits intermediate feature data, as seen in $\mathcal{T}_\text{SL}$. This makes SL advantageous in client-constrained scenarios but introduces higher communication cost due to the transfer of intermediate activations and repeated parameter updates. Overall, while convex methods emphasize theoretical tractability, FL and SL enable practical distributed training by balancing computation and communication in heterogeneous environments.

To reduce the computational complexity associated with ILAC applications, various practical methods have been employed. Notably, model pruning~\cite{9252948} and quantization~\cite{lang2023joint,gu2025task} are two effective methods that substantially reduce the number of required operations during model inference. In particular, model pruning eliminates less significant weights and connections from the network, while quantization reduces the precision of the model’s weights and activations. In practice, these techniques are especially important for enabling the deployment of complex models in real-time wireless communication applications. 

\subsection{Communication Overhead}
Communication overhead arises from the frequent exchange of signals, model parameters, or gradients among distributed nodes in the ILAC framework. It can be quantified based on two key factors, the total volume of data transmitted and the frequency of communication rounds. These factors significantly impact system efficiency, particularly in bandwidth-constrained environments.

In the context of model-distributed designs, such as SL, communication overhead stems from the transfer of intermediate ``smashed data" and partial model parameters between the client and the server. The total communication cost can be expressed as
\begin{align}\label{eq:communication_overhead}
\mathcal{C} = M S_p,
\end{align}
where $M$ is the number of communication rounds and $S_p$ is the size of the transmitted data per round. In SL, $S_p$ depends on the size of the smashed data and local model segments, given by $S_p = 2(dq + N\beta s)$. In contrast, for data-distributed designs such as FL, communication overhead is dominated by the exchange of full model parameters. In this case, $S_p = 2Ns$, reflecting bidirectional transmission across clients.

Note that $S_p$ also characterizes the communication overhead associated with the transmitted signals during inference. Given the substantial challenges  involved in data transmission, numerous techniques have been developed to reduce $S_p$. Specifically, gradient sparsification \cite{lin2023joint} involves transmitting only the most significant gradients, thereby reducing the data volume. 
Also, it is beneficial to employ joint source and channel coding~\cite{8723589} to further compress transmission data, significantly lowering communication costs.

When employing large AI models, the communication cost becomes even more substantial mainly due to the high dimensionality and volume of the parameter matrices. Despite the growing communication burden, large AI models are increasingly desirable in ILAC systems because their strong capabilities in data compression and complex pattern modeling help offset these substantial communication costs. The ability of large models to learn more sophisticated representations can lead to significant reductions in the required transmitted data, especially when applied to tasks that benefit from higher-order abstractions, thus enabling more efficient bandwidth utilization in practice.

\section{Large AI Model and HDC Enhanced ILAC}\label{Sec:lhc}

In this section, we introduce enhancement techniques for ILAC, focusing on the integration of large AI models and HDC. In particular, large AI models, leveraging their powerful generative and generalization capabilities, have significantly improved both performance and efficiency in communication systems\cite{10579546, shen2024Edge,jiang2023large,Bariah2024Telecom}. Meanwhile, HDC, recognized for its lightweight and robust properties, offers an effective solution for various learning- and communication-oriented tasks. We first examine how these technologies individually contribute to ILAC. Then, we further introduce a novel ILAC framework, presenting a case study that effectively integrates large AI models and HDC, optimizing the synergy between learning and communication.  

\subsection{Large AI Model-Enhanced ILAC}
Over the past decades, rapid advancements in artificial intelligence-generated content (AIGC) have driven transformative developments across diverse sectors~\cite{zhao2023survey,xu2024unleashing}. As a core enabler of AIGC, large AI models encompass a broad range of DL architectures. This section explores the emerging applications of large AI models in both model distribution and data distribution frameworks. The integration of these advanced AI models into wireless communication systems holds significant promise for enhancing communication efficiency, facilitating swift decision-making, and enabling new intelligent functionalities. 

One typical category of such integration is physical layer designs leveraging large language models (LLMs)~\cite{zeng2024csigptintegratinggenerativepretrained, 9679803,shao2024wirelessllmempoweringlargelanguage}. In this design paradigm, large pre-trained models are deployed at both the transmitter and receiver ends for effective online inference. A key characteristic of this approach is its centralized nature, where models are initially pre-trained offline and their parameters are then deployed to both user equipment and the BS for real-time communication tasks. This approach leverages the powerful capabilities of LLMs to perform advanced communication tasks such as channel prediction~\cite{zeng2024csigptintegratinggenerativepretrained} and semantic-aware communication~\cite{9679803,xu2025generative}, but also faces challenges related to resource limitations at edge devices.

Despite the promise of LLM-based designs, significant obstacles persist when deploying these architectures on edge devices. In general, the increasing size of these models and the associated computational overhead, particularly during the forward and backward passes, necessitate substantial memory and processing power. For example, LLMs are typically fine-tuned on high-performance GPUs with 40GB or 80GB of memory, requiring over a day of processing time. Consequently, even with efficient tuning methods, adapting relatively smaller models, such as LLaMA-7B\cite{touvron2023llamaopenefficientfoundation}, for deploying on edge devices with limited computational and communication resources remains impractical, especially in 6G applications such as autonomous driving~\cite{9779322} and remote AR/VR collaboration~\cite{10579546}, which demand ultra-low latency and high bandwidth for real-time decision-making and large-scale data transmission.

To mitigate these challenges, various distributed learning techniques have been proposed, particularly emphasizing model distribution and data distribution strategies.

\subsubsection{Model Distribution}
Model distribution involves partitioning a large model across multiple devices, allowing for parallel computation and reducing the computational load for each individual device. Specifically, several studies have introduced SL to distribute LLMs across different devices to achieve an optimal balance between computational complexity and inference efficiency. For example, in \cite{chen2024adaptivelayersplittingwireless}, a model-based reinforcement learning-inspired framework was introduced to dynamically determine optimal model splitting points between the edge and UE. The proposed approach, tested with LLaMA-7B, demonstrated its effectiveness in balancing inference performance with computational load under dynamic network conditions, particularly with respect to varying packet loss rates at the splitting points. Similarly, an SL scheme for fine-tuning LLMs in wireless networks was proposed in \cite{zhang2025splitllmhierarchicalsplitlearning}, where pretrained models and LoRA adapters were strategically split between the cloud, edge servers, and client devices. In this architecture, the edge servers and clients perform parallel training by updating only the LoRA adapters, which are then aggregated in the cloud. This approach achieved a reduction in peak memory usage by up to $74\%$ compared to traditional benchmarks. Moreover, model distribution is also applied in semantic communication scenarios. For example, in \cite{lee2024integrating}, the authors employed BART-base, a pre-trained language model with 1.3B parameters. In this approach, the model was partitioned across different devices, with the semantic encoder deployed on edge servers to handle the heavy computational load of encoding information into discrete indices, while the decoder, requiring fewer computational resources, runs on client devices. The receiver then reconstructs sentences based on the shared codebooks.

\subsubsection{Data Distribution}
Data distribution focuses on allocating data-processing tasks across multiple devices or nodes to facilitate decentralized learning, thereby enhancing communication efficiency and accuracy. Several studies have explored this approach, focusing on reducing communication overhead and improving model efficiency in distributed settings. One common strategy involves FL combined with LLMs. For instance, in \cite{zeng2024csigptintegratinggenerativepretrained}, the authors designed a Swin Transformer-based channel acquisition network integrated with generative pre-training transformer (GPT) for downlink CSI acquisition via federated tuning, which significantly reduces communication overhead by updating only a small portion of the model parameters over the air. Another approach in \cite{Zhao2024FedsLLMFS} proposed combining LoRA with SFL to reduce computational load by splitting the model between clients and servers. Additionally some research has focused on combining LLMs with MARL. Specifically, the study in \cite{10638533} introduced CommLLM, a multi-agent system designed for communication tasks exploiting natural language. In particular, the system incorporated multi-agent data retrieval, collaborative planning, and evaluation to solve communication tasks efficiently, validated through a case study in a 6G semantic communication system. Moreover, in \cite{10582829}, a pre-trained LLM was employed for channel prediction to predict future downlink CSI based on historical uplink CSI data. The model leveraged GPT-2 as its backbone, applying fine-tuning while keeping most parameters frozen, thereby achieving efficient channel prediction performance.

In summary, the integration of large AI models into wireless communication systems through model and data distribution strategies offers significant potential to enhance communication efficiency of ILAC. Specifically, model distribution facilitates parallel computation, alleviating the computational load on edge devices, while data distribution methods, such as FL and multi-agent systems, enable decentralized learning and reduce communication overhead. These advancements are crucial for adapting large-scale AI models to resource-constrained devices in 6G networks. To fully harness these technologies, establishing an ILAC framework is essential, enabling seamless interactions between AI models and communication systems, jointly optimizing computational and communication resources, improving the scalability, efficiency, and adaptability of next-generation wireless networks, and paving the way for more intelligent, and data-driven communication systems.

Despite the benefits of model and data distribution strategies, both approaches require resource-intensive fine-tuning processes~\cite{9357490, lin2024efficient, chen2024communication}. As a computationally efficient alternative, HDC emerges as a promising method to reduce computational overhead. Consequently, we propose the HDC-enhanced ILAC framework, aiming to further optimize performance in wireless communication systems.
\subsection{Hyperdimensional Computing Enhanced ILAC}




HDC offers a lightweight and robust solution to effectively integrate large AI models into ILAC, substantially reducing learning overhead while preserving the strong generalization strength of large models. By encoding data into high-dimensional vector representations \cite{Kleyko2022SurveyHyperdimensional,Kleyko2023SurveyHyperdimensional}, HDC significantly reduces the computational burden associated with both model training and training processes. Furthermore, HDC enhances system robustness by distributing information across numerous dimensions, thereby improving noise resistance and enabling efficient parallel processing. This capability is particularly beneficial for resource-constrained environments, such as wireless networks. Additionally, HDC facilitates rapid model adaptation to diverse tasks without the need for extensive retraining, perfectly aligning with ILAC’s need for dynamic and intelligent decision-making. Consequently, by integrating HDC, ILAC not only alleviates the computational challenges posed by large AI models but also enhances its ability to support next-generation intelligent communication networks. In the following, we provide the implementation details of HDC.

HDC is a computational paradigm inspired by human brain's  information processing mechanisms. 
The fundamental concept of HDC leverages the unique properties of extremely high-dimensional and random patterns by employing hypervectors (HVs), i.e., very high-dimensional random vectors, as the basis for computation. By operating in such high-dimensional spaces, HDC enables significant parallelism, facilitating rapid data processing for both training and inference. Indeed,
the high-dimensional nature of HDC endows it with robustness and inherent orthogonality~\cite{Chang_Chuang_Huang_Wu_2023}. In practice, HVs are resilient to variations in individual bits since the overall vector pattern and structure determine their information content, rendering HDC highly resistant to noise and interference~\cite{Kleyko2022SurveyHyperdimensional}. Furthermore, owing to their large dimensionality, typically tens of thousands of dimensions, each HV tends to be nearly orthogonal to others, ensuring a large pool of usable vectors from which selections can be made with only minimal overlap~\cite{basaklar2021hypervector}. This near-orthogonality typically arises from randomly generated patterns used to construct models. In fact, this advantage of randomness lies in the fact that meaningful vectors exhibit large Hamming distances, resulting in very low cosine similarity between two HVs. When meaningful vectors are represented as 10,000-bit vectors, even if noise corrupts more than one-third of the bits, the resulting vector can still be accurately matched to the correct vector~\cite{kanerva2009hyperdimensional,basaklar2021hypervector}.

In HDC, the basic operations on HVs include addition, multiplication, and similarity measurement. Specifically, addition serves as a bundling mechanism, i.e., for aggregating multiple HVs, while multiplication, typically executed as an element-wise XOR, is exploited for binding them together. Moreover, similarity measures, such as Hamming distance and cosine similarity, quantify the resemblance between HVs, ensuring accurate retrieval and processing. These operations enable HDC to leverage the high-dimensionality and randomness of HVs in unique ways, allowing them to represent sets and bind variables to values effectively. 

Specifically, the summation of HVs represents a set and the product of HVs binds variables to values. For example, if the HV for a feature $x$ is denoted by $X$ and the HV for its corresponding value $a$ is denoted by $A$, they can be bound together through multiplication, resulting in the binding $X \times A$. During unbinding, the original value can be retrieved by multiplying the binding with the feature's HV. To bundle multiple records, addition is employed to sum the records, representing them as a set. For instance, $H = X \times A + Y \times B + Z \times C$ denotes an aggregation of three different records. Notably, the sum of HVs retains a high degree of similarity to each individual HV being aggregated. In contrast, the product of two HVs is nearly orthogonal to the original input HVs \cite{kanerva2009hyperdimensional}.
Given a set of $n$ hypervectors, the computational complexity of binding, bundling, and permutation operations in HDC is $\mathcal{O}(nd)$, where $d$ represents the dimensionality of the hypervectors.


\begin{figure}[t]
    \centering
    \includegraphics[width=\linewidth]{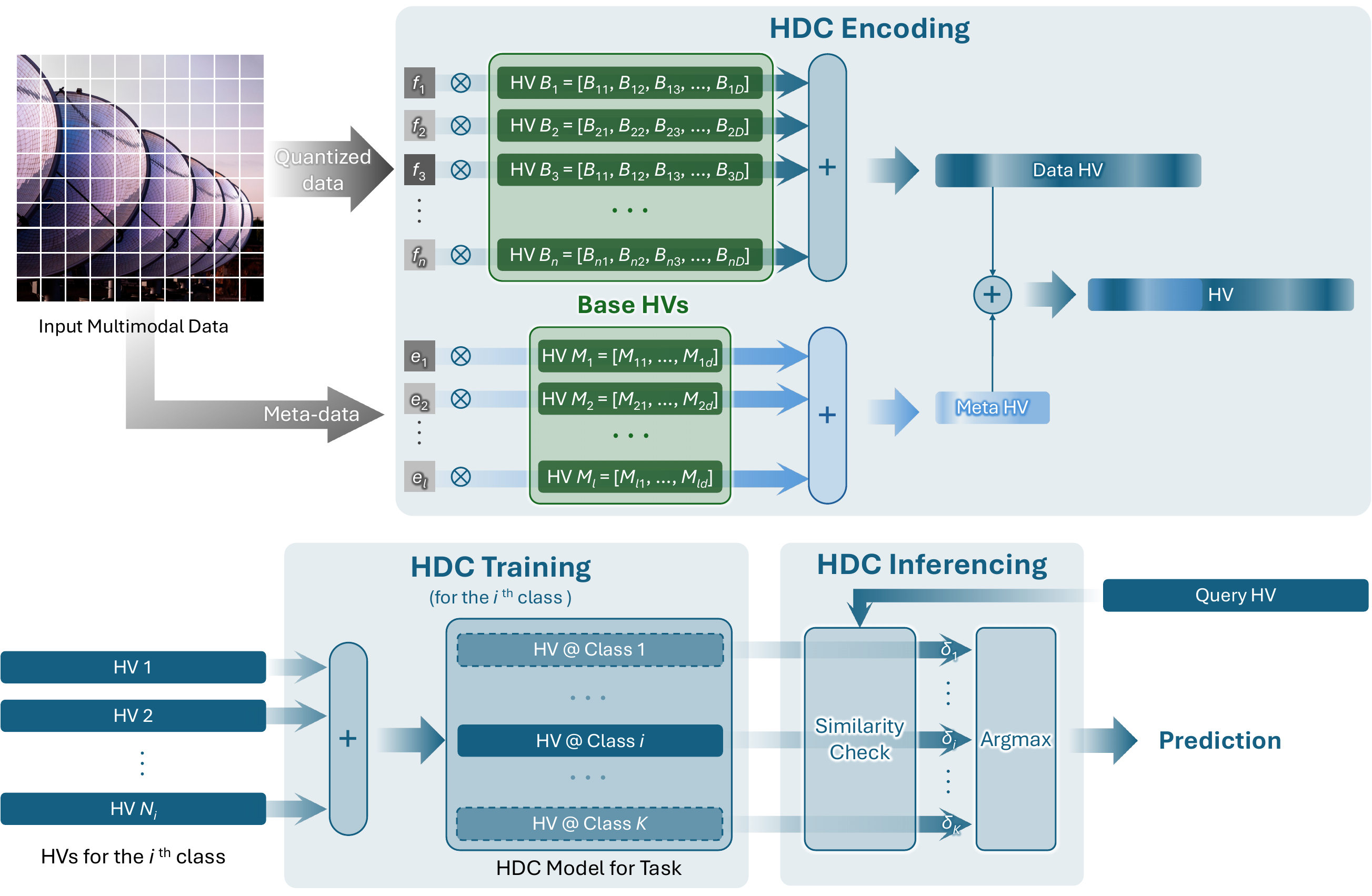}
    \caption{A workflow of HDC encoding, training, and inferencing.}
    \label{fig:HDC}
\end{figure}


Fig.~\ref{fig:HDC} depicts the primary workflow of HDC,  encompassing encoding, training, and inferencing. The input multimodal data consists of $n$ pixel values $\{f_1, \dots, f_n\}$. This data is encoded into a hypervector, termed Data HV, exploiting a random projection method:
\begin{equation}\label{eq:dataHV}
    \text{Data HV} = B_1\times f_1+B_2\times f_2+\dots+B_n\times f_n,
\end{equation}
where $\{B_1, \dots, B_n\}$ are the corresponding base HVs, with each $B_i\in \{+1,-1\}^D$.
To further encode useful metadata $\{e_1, \dots, e_l\}$, where $l$ represents the number of useful metadata items (such as data type, bit depth, and format), the method in \cite{imani2019framework} injects the metadata into small segments of the encoded hypervector to retain this information. This yields Meta HV, expressed as
\begin{equation}\label{eq:metaHV}
    \text{Meta HV} = M_1\times e_1+M_2\times e_2+\dots+M_l\times e_l,
\end{equation}
where $\{M_1, \dots, M_l\}$ are the corresponding base meta hypervector, and $M_i\in \{+1,-1\}^d,$ with $d<D$. The metadata injection procedure is given by
\begin{equation}\label{eq:HVall}
    \text{HV} = \text{Data HV} + \{\text{Meta HV}, 0,\dots, 0\}.
\end{equation}

As previously mentioned, we can represent sets leveraging the addition of HVs. In the training phase, HDC leverages this property to encapsulate entire classes in a single high-dimensional representation. For a classification problem with $K$ classes, each class $ k $ is represented by a class HV $ \text{AM}_k $, obtained by,
     \begin{equation}\label{eq:AM}
     \text{AM}_k = \sum_{i=1}^{N_k} \text{HV}_{i},
     \end{equation}
where $ N_k $ is the number of samples in class $ k $ and $\text{HV}_i$ represents the HV corresponding to the $i$-th sample of class $k$. This single-pass summation process effectively captures collective information, eliminating the need for iterative optimization via backpropagation required in neural networks \cite{Kleyko2022SurveyHyperdimensional, Kleyko2023SurveyHyperdimensional}.

During the inference phase, the class of a query HV, denoted by $ \text{HV}_{\text{query}} $, is determined by measuring its similarity to each class HV $ \text{AM}_k $. The similarity check can be conducted using either cosine similarity or Hamming distance as
     \begin{equation}
     \delta_k = \text{similarity}(\text{HV}_{\text{query}}, \text{AM}_k).
     \end{equation}
Subsequently, the predicted class is the one with the highest similarity score that is given by
     \begin{equation}
     \text{prediction} = \arg\max_k \delta_k.
     \end{equation}
In the similarity calculation process for classification with $K$ classes, the complexity is $\mathcal{O}(K d)$.


Several reviews on HDC already exist \cite{kanerva2009hyperdimensional, Kleyko2022SurveyHyperdimensional, Kleyko2023SurveyHyperdimensional}, primarily focusing on hardware architectures and natural language processing. 
However, HDC offers a particularly promising approach for integrating learning and communication. Its efficiency, robustness, and scalability make it highly suitable for various applications, from enhancing wireless communication to ensuring privacy and facilitating collaborative learning. As HDC continues to evolve, its potential to revolutionize communication systems and intelligent applications is expected to grow.
Now, we discuss the current applications of HDC in ILAC systems from the perspectives of communication-oriented and learning-oriented approaches, respectively.

\subsubsection{Communication-oriented}

In HDC, the inherent high-dimensional characteristics of HVs provide robustness against noise, making them particularly suitable for processes such as compression, encoding, and modulation in communication tasks.

In practice, HDC can be leveraged to overcome the challenges of reliable and efficient wireless communication in noisy and high-interference environments, as well as to develop novel data transmission methods that enhance communication performance in dense and adverse wireless conditions. For example, multiple wireless sensing devices simultaneously transmit HVs representing their sensed information. The receiver then analyzes these superimposed HVs by summing those from each sensing device and comparing the result against a predefined threshold. This approach eliminates the need for multiple access schemes, making it well-suited for scenarios in which multiple devices must report their status to a central node. In \cite{jakimovski2012collective}, multiple devices transmitted HVs associated with temperature data, enabling the receiver to detect specific temperature conditions or determine how many devices were exposed to a particular temperature by analyzing the superimposed HVs. Similarly, in \cite{kleyko2012dependable}, sensing devices constructed composite HVs by applying multiplicative binding and superposition to represent their sensed data, and the receiver recovered the information represented in the received composite HVs by exploiting the base HVs, demonstrating HDC's efficiency and robustness in complex communication scenarios.

HDC has also been successfully applied to integrate channel coding and modulation in communication systems, referred to as hyperdimensional modulation (HDM) \cite{kim2018hdm,hersche2021near,hsu2023hyper}. In HDM, complex-valued HVs represent individual data segments, and multiple information bits are transmitted by combining several approximately orthogonal HVs. This strategy distributes information bits across multiple elements of the HV, enabling HDM to exhibit high resilience against element-level failures, especially in noise-dominated communication environments. As 6G systems are expected to support ultra-reliable low-latency communication (URLLC) and massive machine-type communication (mMTC) with highly dynamic and noisy channels, the resilience of HDM makes it a suitable solution for such environments. 
Furthermore, the iterative decoding of the received composite HVs at the receiver can significantly improve the code rate, making it adaptable to the high throughput and low latency demands of 6G networks.
For instance, in \cite{kim2018hdm}, a soft feedback iterative decoding scheme based on continuous interference cancellation was employed for HDM demodulation, enhancing the system's ability to handle interference, a critical aspect in 6G networks. Moreover, in \cite{hersche2021near}, HDM was shown to rely on the complex-valued components of HVs to ensure a bit error rate (BER) performance comparable to that of traditional low-density parity-check (LDPC) and polar codes, and it was also adopted for near-channel classification, enabling efficient channel management. In addition, HDM was utilized to ensure robust communication of short packets in mMTC in \cite{hsu2023hyper}, where a CRC-aided $K$-best decoding algorithm was proposed to ensure a low packet error rate (PER), which is essential for fulfilling  reliability requirements.

Another advantage of HDC, particularly in edge computing and IoT, is its unique capability to simultaneously handle multiple cognitive tasks, known as multi-task learning (MTL). As 6G systems are expected to support diverse and resource-constrained applications, the ability of HDC to perform MTL efficiently aligns with the multi-functional demands of 6G.
In \cite{chang2021multa}, MulTa-HDC was proposed as an MTL framework for HDC that balances memory overhead and performance by integrating shared and independent associative memory (AM) approaches. The dimension ranking for effective associative memory sharing (DREAMS) merges shared AMs while preserving task-specific information, and dimension ranking for independent memory retrieval (DRIMER) mitigates interference in independent AMs, providing an efficient and scalable solution for the complex multi-tasking requirements in 6G.

\subsubsection{Learning-oriented}
Applications and the potential of HDC in fields such as FL and edge computing were demonstrated in \cite{imani2019framework,hsieh2021fl,zhao2022fedhd,kasyap2023hdfl}, highlighting notable improvements in privacy protection, computational efficiency, and model robustness.

In edge computing scenarios, clients may lack sufficient channel bandwidth to transmit individual HVs for each data item. However, in FL, each client can independently train an initial model leveraging its own data, with each class represented by a single HV. Once the edge server receives these initial models from all clients, it  aggregates them to establish a global model. For example, in \cite{imani2019framework}, a reliable HDC-based collaborative learning framework was introduced. Specifically, this framework first creates distinct key HVs for each client and the cloud server based on the Multi-Party Computation (MPC) protocol. These client-specific key HVs are exploited for data encoding to preserve user privacy, ensuring that the data encoded with key HVs cannot be directly interpreted or accessed by the cloud server. The cloud server can still perform model aggregation based on the encoded data through permutation, but without revealing sensitive user data. Similarly, in \cite{hsieh2021fl},  FL-HDC was proposed to enhance the system's ability to handle heterogeneous data and cope with network instability. Moreover, in \cite{zhao2022fedhd}, FedHD was introduced to improve the efficiency and accuracy of model training by combining FL and HDC. Furthermore, in \cite{kasyap2023hdfl}, HDFL was presented to integrate privacy-preserving HDC with a robust FL framework, demonstrating strong performance in handling malicious attacks and managing noisy data.

\subsection{Case Study: A New Framework for ILAC}
\begin{figure*}[t]
    \centering
    \includegraphics[width=.9\linewidth]{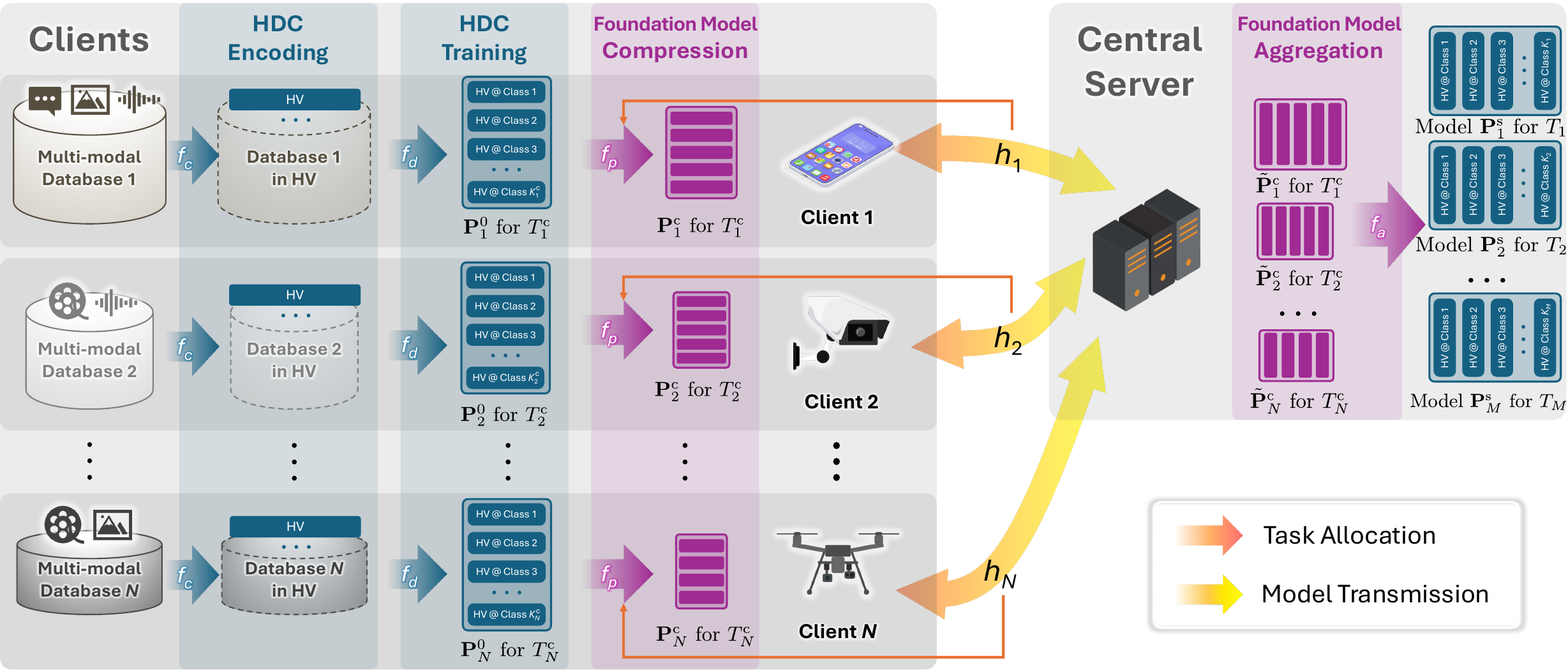}
    \caption{The novel framework for effective ILAC encompasses HDC encoding, local training, and compression via a foundation model at each client device. The compressed models are then transmitted to a central server for aggregation, resulting in the global model update.  In this context, $C_j$ represents the $j$-th client, $T_i$ denotes the $i$-th task, and $K_i$ indicates the number of classes associated with task $T_i$. Specifically, $T_j^\text{c}$ and $K_j^\text{c}$ correspond to the set of tasks and their respective class numbers handled by client $C_j$.}
    \label{fig:new_framework}
\end{figure*}


To concretely realize the general ILAC optimization problem \eqref{eq.gop} formulated in Section \ref{Sec:fi}, we build a task-specific implementation that integrates large AI models and HDC. In traditional deployments, adapting a foundation model to downstream communication tasks often requires task-specific fine-tuning, which introduces significant training overhead due to the need for gradient-based backpropagation. To effectively address this challenge, we propose to leverage HDC as a lightweight yet versatile alternative to task-specific fine-tuning networks. HDC models, recognized for their inherent task generalizability and training-free nature, allow us to bypass the need for gradient-based updates while retaining sufficient adaptability to diverse learning objectives in  communication scenarios~\cite{Chang_Chuang_Huang_Wu_2023}.

Building upon this foundation, we construct an ILAC framework that jointly considers computational cost, communication performance, and inference accuracy. These factors are integrated into a unified optimization problem, echoing the general structure described by problem \eqref{eq.gop}. Through this case study, we aim to quantitatively demonstrate the advantages of the ILAC approach over single function designs that are either purely communication-oriented or purely learning-oriented. The following detailed case study provides important insights into this framework and its performance across representative scenarios.

\subsubsection{System model}

We consider a distributed system, where $ M $ classification tasks, each involving $ K_i $ classes ($i = 1,2,\dots, M$), are allocated to $ N $ heterogeneous clients ($N\ge 2M$), such as cameras, smartphones, and UAVs, c.f. Fig.~\ref{fig:new_framework}. Each client independently trains a classification model for the assigned task by employing HDC. The trained HDC models, represented by multiple HVs, are then compressed by a foundation model before being transmitted to a central server, where they are processed and aggregated using another foundation model.

\paragraph{Task assignment}
Let $ \mathcal{T} = \{T_1, T_2, \ldots, T_M\} $ denote the set of classification tasks and let $ \mathcal{C} = \{C_1, C_2, \ldots, C_N\} $ represent the set of clients. The task assignment can be represented by a binary matrix $ \mathbf{A} \in \{0,1\}^{M \times N} $, where $ A_{ij} = 1 $ if task $ T_i $ is assigned to client $ C_j $ and otherwise $ A_{ij} = 0 $.  Moreover, to reduce sensitivity to noise and outliers and enhance the robustness of the class representation, we assume that each task $ T_i $ is assigned to at least two clients:
\begin{equation}
 \sum_{j=1}^{N} A_{ij} \geq 2, \quad \forall i.
\end{equation}
Additionally, $T^\text{c}_{j} =\sum_{i=1}^M A_{ij}T_i $ and $K^\text{c}_{j} =\sum_{i=1}^M A_{ij}K_i $ represent a task assigned to client $ C_j $ and the number of classes handled by client $ C_j $, respectively. 

\paragraph{Client training}

Each client $ C_j $ adopts its local multi-modal database to train an HDC model for its assigned task $T^\text{c}_{j}$. This training process includes data encoding via an HDC encoder $ f_c(\cdot) $ and model training through an HDC trainer $ f_d(\cdot) $. It follows
\begin{equation}
\mathbf{P}^0_j = f_d(f_c(\mathbf{D}_j)),
\end{equation}
where $\mathbf{P}^0_j$ and $\mathbf{D}_j $ represent the obtained HDC model and the multi-modal database at client $ C_j $. The size of the HDC model is given by 
\begin{equation}
    s^0_j=K^\text{c}_j D=\sum_{i=1}^M A_{ij}K_i D,
\end{equation}
where $K_i$ is the number of task classifications and $D$ is the dimension of the hypervectors.

\paragraph{Model compression}
Given the channel state $h_j\in\mathbb C$, the HDC models $\mathbf{P}^0_j$ are preprocessed exploiting a foundation model to improve the transmission efficiency, i.e.,
\begin{equation}
\mathbf{P}^\text{c}_j = f_p(\mathbf{P}^0_j, h_j),
\end{equation}
where $f_p(\cdot)$ is the foundation model-based preprocessing function, which can include techniques such as compression or fragmentation based on the channel conditions.

\paragraph{Model transmission}
After training and compression, each client $C_j$ sends the compressed model $ \mathbf{P}_j $ to  the central server.
In the uplink transmission, the model received at the central server is expressed as
\begin{equation}
    \tilde{\mathbf P}^\text{c}_j = \mathbb{P} (\mathbf{P}^\text{c}_j \mid  h_{j}),
\end{equation}
where $\mathbb{P} (\mathbf{P}^\text{c}_j \mid h_j)$ represents the conditional probability of the uplink model $\tilde{\mathbf{P}}^\text{c}_j$ at the central server, given the channel state information $h_j$.
In this process, the transmission time $ t_j $ for client $ C_j $ depends on the compressed model size $s^\text{c}_j$ and the transmission rate $ r_j $, and
\begin{equation}
t_j = \frac{s^\text{c}_j}{r_j},
\end{equation}
where the transmission rate is given by
\begin{equation}\label{eq:commurate}
r_j = b_j \log_2 \left(1 + \frac{p_j |h_j|}{\sigma^2 b_j}\right),
\end{equation}
where $b_j$ is the bandwidth, $p_j$ is the transmission power, $|h_j|$ is the channel gain, and $\sigma^2$ is the noise power.

\paragraph{Server-side processing and model aggregation}

The central server receives the HDC models from $N$ clients and processes each received model, $\tilde{\mathbf P}^\text{c}_j$, using a foundation model.  An aggregation function, $ f_a(\cdot) $, combines these processed models into a unified model $ \mathbf{P}^\text{s}_i $ for each task $ T_i $ as
\begin{equation}
\mathbf{P}^\text{s}_i = f_a(\{\tilde{\mathbf P}^\text{c}_j \mid A_{ij} = 1\}).
\end{equation}
During aggregation, both the contribution of each client and the accuracy of the respective models are taken into account.


\subsubsection{Problem formulation}


We formulate a concrete optimization problem that builds on generalized optimization problem~\eqref{eq.gop}. The main goal of this optimization is to improve both communication and learning performance while respecting resource constraints. In this context, we specifically instantiate the abstract communication and learning objectives from the generalized problem to reflect the key performance metrics of our system.

For the communication-related objective $f(\cdot)$, we consider the communication rate in \eqref{eq:commurate} as the primary metric, representing the efficiency of data transmission. For the learning-related objective $g(\cdot)$, we focus on two critical aspects: the computational cost caused by compressing the model, which indicates the level of model size reduction, and the inference accuracy, which reflects the effectiveness of the model in producing correct results. Specifically, the computational cost for each client $C_j$ is related to the compression ratio, expressed as
\begin{equation}\label{eq:ej_computational_cost}
    e^\text{c}_j =\eta \left(\frac{s^0_j}{s^\text{c}_j}\right), \quad \forall j,
\end{equation}
where $ \eta(\cdot) $ characterizes the  processing capability of the client-side foundation model \cite{basaklar2021hypervector}.
The performance for each task $T_i$ is primarily related to the accuracy of each participating aggregated model and the communication environment, as given by
\begin{equation}\label{eq:qi_performance}
    Q = \underbrace{\sum_{i=1}^M \sum_{j=1}^N A_{ij}\, r_j}_\text{Communication performance} \times \underbrace{\frac1M \sum_{i=1}^M \frac{\sum_{j=1}^N A_{ij}\phi(\tilde{\mathbf{P}}^\text{c}_j)}{\sum_{j=1}^N A_{ij}}}_\text{HDC performance} ,
\end{equation}
where $ \phi(\cdot) $ measures the accuracy of the HDC model and can be determined by the accuracy on the validation set.

To integrate these objectives, we define a unified performance evaluation metric known as the \textit{Cost-to-Performance Ratio (CPR)}, which quantifies the trade-off between the computational cost and the overall system performance in terms of learning and communication efficiency~\cite{qian2023user}. In terms of the optimization variables, the communication-related variables include bandwidth and power allocations, while the learning-related variables involve task assignments and model size.

The minimization problem involves optimizing task assignment $\mathbf A$, model size $\mathbf s=(s^\text{c}_j)|_{j\in\mathcal C}$, bandwidth $\mathbf b = (b_j)|_{j\in\mathcal C}$, and transmission power $\mathbf p = (p_j)|_{j\in\mathcal C}$. The minimization problem can be mathematically formulated as
\begin{subequations}\label{prob:framework}
    \begin{alignat}{1}
\min_{\mathbf A, \mathbf s, \mathbf b, \mathbf p}\quad&\frac{\sum_{j=1}^N e^\text{c}_j}{Q}\tag{\ref{prob:framework}}\\ 
\text{s.t.}\quad
& \sum_{j=1}^{N} A_{ij} \geq 2, \quad \forall i, \label{eq:framework_a1}\\
& \sum_{i=1}^{M} A_{ij} = 1, \quad \forall j, \label{eq:framework_a2}\\
& A_{ij}\in\{0,1\},\quad\forall i,j, \label{eq:framework_a3}\\
& s^\text{c}_j \le \sum_{i=1}^M A_{ij}K_i D,\quad \forall j, \label{eq:framework_s}\\
& e^\text{c}_j\le e^\text{c}_{\max,j}, \quad \forall j, \label{eq:framework_e}\\ 
&\sum_{j=1}^{N}b_j \le b_{\max}, \label{eq:framework_d1}\\ 
&p_j \le p_{\max,j}, \quad \forall j,  \label{eq:framework_d2}\\
& b_j\ge 0, p_j\ge 0,\quad \forall j, \label{eq:framework_d3}\\
& t_j \le t_{\max},\quad \forall j, \label{eq:framework_t}
    \end{alignat}
\end{subequations}
where $ e^\text{c}_j $ and   $ Q $ represent the computational costs for client $ C_j $ and  the performance for tasks, defined in \eqref{eq:ej_computational_cost} and \eqref{eq:qi_performance}, respectively. Constraints \eqref{eq:framework_a1}, \eqref{eq:framework_a2}, and \eqref{eq:framework_a3} indicate that each task must be assigned to at least two clients, and each client can only handle one task. Constraint \eqref{eq:framework_s} considers the process of compression using a foundation model. Constraint \eqref{eq:framework_e} represents the maximum computational cost constraint for each client, where $ e^\text{c}_{\max,j} $ is the maximum computational cost that client $ C_j $ can handle.  Constraints \eqref{eq:framework_d1}, \eqref{eq:framework_d2}, and \eqref{eq:framework_d3} limit the  constraints on bandwidth and transmission power for clients, where $ b_{\max} $ and $p_{\max,j}$ are the maximum bandwidth and transmission power that the system can handle, respectively.  Constraint \eqref{eq:framework_t} limits the transmission latency cost, where $ t_{\max} $ is the maximum tolerable transmission latency from the client to the server.

\begin{figure*}[t]
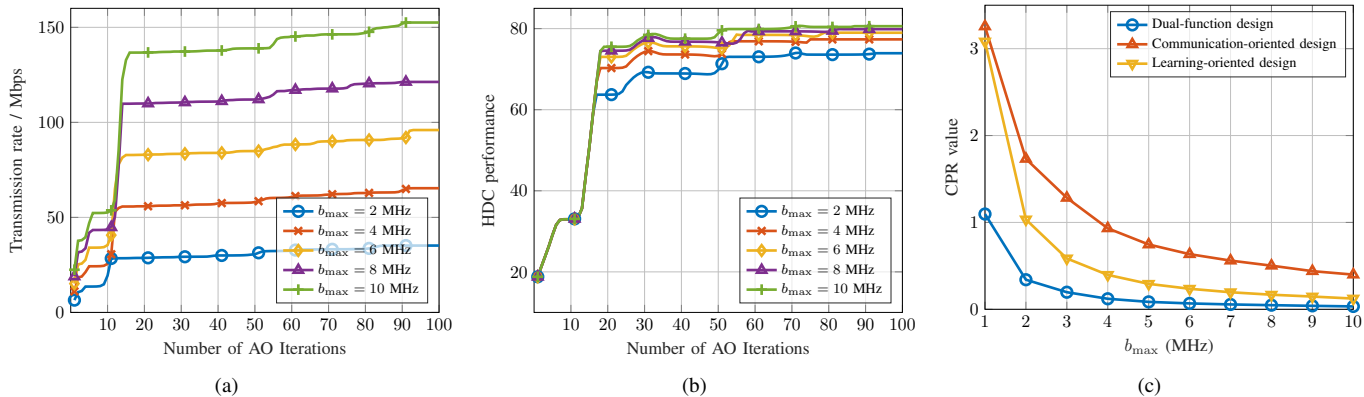

    \subfigure[]{\input{fig/Section5/iter_Q_Comm}\label{fig:iter_Q_Comm}}\hfill
    \subfigure[]{\input{fig/Section5/iter_Q_HDC}\label{fig:iter_Q_HDC}}\hfill
    \subfigure[]{
%
\definecolor{mycolor1}{rgb}{0.00000,0.44700,0.74100}%
\definecolor{mycolor2}{rgb}{0.85000,0.32500,0.09800}%
\definecolor{mycolor3}{rgb}{0.92900,0.69400,0.12500}%
\begin{tikzpicture}[scale=.67]

\begin{axis}[%
width=3.5in,
height=3in,
xmin=1,
xmax=10,
xlabel style={font=\color{white!15!black}},
xlabel={$b_{\max}$ (MHz)},
ymin=0,
ymax=3.5,
ylabel style={font=\color{white!15!black}},
ylabel={CPR value },
axis background/.style={fill=white},
xmajorgrids,
ymajorgrids,
yminorgrids,
xtick={1,2,3,4,5,6,7,8,9,10},
legend style={legend cell align=left, align=left, draw=white!15!black,fill opacity=0.7,text opacity=1,font=\footnotesize}
]
\addplot [color=mycolor1, line width=1.5pt, mark size=3pt,  mark=o, mark options={solid, mycolor1}]
  table[row sep=crcr]{%
1	1.0965830502467\\
2	0.340464975172087\\
3	0.19669555276919\\
4	0.121233660251777\\
5	0.0854748757675917\\
6	0.0676782676096602\\
7	0.0560051949093318\\
8	0.0476183664688862\\
9	0.0399080808849852\\
10	0.0331417049488281\\
};
\addlegendentry{Dual-function design}

\addplot [color=mycolor2, line width=1.5pt, mark size=3pt, mark=triangle, mark options={solid, mycolor2}]
  table[row sep=crcr]{%
1	3.25617958185503\\
2	1.73163851969551\\
3	1.28290840112597\\
4	0.932078083596546\\
5	0.746047920945622\\
6	0.634736570868906\\
7	0.560668944602183\\
8	0.50226695065558\\
9	0.439294435701732\\
10	0.399352934744659\\
};
\addlegendentry{Communication-oriented design}

\addplot [color=mycolor3, line width=1.5pt, mark size=3pt, mark=triangle, mark options={solid, mycolor3, rotate=180}]
  table[row sep=crcr]{%
1	3.08188330074036\\
2	1.03405876343086\\
3	0.585291164163994\\
4	0.396033284437019\\
5	0.292875938562504\\
6	0.236337437639165\\
7	0.196291914712638\\
8	0.167866093887025\\
9	0.146736314546679\\
10	0.123477539049159\\
};
\addlegendentry{Learning-oriented design}

+\end{axis}
\end{tikzpicture}%
\label{fig:bmax_f}}
        \caption{
(a) Transmission rate vs. number of iterations for different $b_{\max}$,
(b) HDC performance vs. number of iterations for different $b_{\max}$, and
(c) CPR vs. $b_{\max}$. }
        \label{fig:simulation_result}
\end{figure*}

\subsubsection{Roadmap of the algorithm}
In this case study, we adopt traditional optimization techniques to evaluate the performance of the proposed ILAC framework. This methodological choice is motivated by the nature of the HDC-based inference approach. Unlike conventional learning models that rely on gradient-based updates, the HDC paradigm operates without gradient information during inference. Consequently, the optimization process is decoupled from standard learning-based methods, rendering machine learning techniques unnecessary in this context.

The use of traditional optimization allows for a clear and tractable evaluation of the ILAC framework in terms of resource allocation and communication efficiency. This approach provides a baseline analysis that highlights the intrinsic capabilities of the framework, without the additional complexity introduced by model training. Moreover, this preliminary investigation serves as a foundational reference for future work, where more sophisticated learning-based strategies may be incorporated to further enhance system performance.

We employ the Dinkelbach algorithm and alternating optimization (AO) to address problem \eqref{prob:framework}. By applying Dinkelbach's transformation and introducing the auxiliary variable $\lambda$, optimization problem \eqref{prob:framework} can be equivalently reformulated as 
\begin{equation}\label{prob:framework_Dinkelbach}
    \begin{split}
    \min_{\mathbf A, \mathbf s, \mathbf b, \mathbf p}\quad&{\sum_{j=1}^N e^\text{c}_j} -\lambda {Q} \\
    \text{s.t.}\quad& \eqref{eq:framework_a1}- \eqref{eq:framework_t}.
    \end{split}
\end{equation}

The Dinkelbach algorithm is then iteratively applied to optimize $\lambda$ and solve problem  \eqref{prob:framework_Dinkelbach}. In the $i$-th iteration, given $\lambda^{(i-1)}$, problem \eqref{prob:framework_Dinkelbach} is solved using the AO method, yielding $\mathbf{A}^{(i)}, \mathbf{s}^{(i)}, \mathbf{b}^{(i)}, \mathbf{p}^{(i)}$. Subsequently, the CPR value $\lambda^{(i)}$ is updated based on the optimized variables. This process is repeated until convergence.


Specifically, we first employ the Dinkelbach algorithm to transform the fractional objective in \eqref{prob:framework} into a parameterized form. Then, we adopt the AO method to iteratively optimize $\{\mathbf{A}, \mathbf{s}\}$ and $\{\mathbf{b}, \mathbf{p}\}$. In the first step of AO, we fix $\{\mathbf{b}, \mathbf{p}\}$ and optimize $\{\mathbf{A}, \mathbf{s}\}$, where the original problem is reformulated as a mixed-integer programming problem and solved using a combination of Integer Linear Programming (ILP) and a greedy allocation strategy. In the second step of AO, $\{\mathbf{A}, \mathbf{s}\}$ are fixed while $\{\mathbf{b}, \mathbf{p}\}$ are optimized by converting the non-convex problem into a convex one using fractional programming methods. Finally, the CPR value $\lambda$ is updated based on the current solutions, and the iterative process continues until convergence.

In each AO iteration, the first step optimizes task assignment and model size, with a complexity of $\mathcal{O}(MN + N^3)$, where $M$ is the number of tasks and $N$ is the number of users. The second step optimizes bandwidth and power allocation through fractional programming, which also takes $\mathcal{O}(N^3)$. Thus, each AO iteration costs $\mathcal{O}(MN + N^3)$. Considering the outer Dinkelbach loop with $I_1$ iterations and the inner AO loop with $I_2$ iterations, the total complexity is $\mathcal{O}(I_1 I_2 (MN + N^3)),$ which scales polynomially with the number of users and tasks and linearly with the the number of iterations. Since these two blocks are optimized in a sequential manner rather than jointly, the complete subproblem is not solved in a single iteration. This implies that, at best, we achieve locally optimal solutions for each individual subproblem. Under this AO scheme, the algorithm ensures monotonic improvements of the energy-efficiency objective in each iteration. However, it only guarantees convergence to a stationary point, which might be suboptimal in a global sense. While empirical results demonstrate that the AO iteration with Dinkelbach algorithm converges reliably, we explicitly acknowledge that the global optimality guarantee is preserved only if each subproblem is solved to exact optimality—a condition that may not always hold in practical implementations.

\subsubsection{Simulation Results}
We examine a network topology spanning an area of 100~m~$\times$~100~m, consisting of 12 heterogeneous clients and 5 distinct classification tasks, each corresponding to a class set from the following: $\{10, 20, \dots, 50\}$. The channel fading coefficient, denoted by $h_j$, between any client $C_j$ and the server is given by $128.1 + 37.6 \log_{10} d_j$, where $d_j$ denotes the Euclidean distance between client $C_j$ and the server~\cite{qian2023user}. The power of Gaussian noise is fixed at $\sigma^2 = -134$~dBm. Each server has a total bandwidth of $b_{\max} \in \{1,2,\dots, 10\}$~MHz, and the maximum transmit power for each client is limited to $p_{\max} = 0.2$~W. The maximum permissible transmission time from any client to the server is limited to $t_{\max} = 0.5$~s.

Fig.~\ref{fig:simulation_result} illustrates the performance of the proposed algorithm in terms of transmission rate, HDC performance, and CPR values for different total bandwidths, providing a comprehensive evaluation of its effectiveness. Two baseline designs are considered: a communication-oriented baseline that optimizes only the transmission rate in \eqref{eq:qi_performance}, and a learning-oriented baseline that focuses solely on enhancing HDC performance. Fig.~\ref{fig:iter_Q_Comm} and Fig.~\ref{fig:iter_Q_HDC} illustrate the convergence of the transmission rate and HDC performance, respectively, with the algorithm stabilizing after approximately $90$ iterations. As the total bandwidth $b_{\max}$ increases, both the transmission rate and HDC performance improve significantly, demonstrating that the proposed algorithm effectively converges to a stationary point of problem \eqref{prob:framework}. 
Fig.~\ref{fig:bmax_f} compares the variation of CPR values of different designs as $b_{\max}$ changes. The CPR value consistently decreases as the bandwidth constraint $b_{\text{max}}$ increases. Moreover, the proposed dual function design, which jointly optimizes communication and learning objectives, achieves lower CPR values for all bandwidths. It consistently outperforms both the communication-oriented and learning-oriented designs, thereby demonstrating the effectiveness of the proposed joint optimization strategy.

\section{Conclusion and Open Problems}\label{Sec:c}

In this paper, we have presented a comprehensive overview of ILAC methodologies and performance optimization. In particular, we have proposed a unified framework and generalized optimization formulation, bridging learning performance and communication efficiency. We further have analyzed various model and data distribution strategies, highlighting their practical implications and challenges. Finally, to address these challenges, we have proposed an enhanced ILAC framework that integrates large AI models for enhanced generalization and HDC for lightweight representation, demonstrating its effectiveness by a case study on cost-to-performance ratio optimization.

Although significant progress has been made with the ILAC framework, several open problems remain that require further investigation.
\subsection{Cross-Layer Resource Orchestration} The coexistence of large AI models and HDC in ILAC introduces significant challenges in resource management spanning the device-edge-cloud continuum. Large AI models, often comprising billions of parameters, impose heavy computational and memory demands, intensifying the computational burden during distributed inference and training. Concurrently, the high-dimensional operations in HDC, involving tens of thousands of dimensions, lead to substantial storage and transmission overhead, complicating real-time resource optimization at the edge. For example, SL partitions models across nodes to reduce local computation but amplifies intermediate feature transmission, creating a trade-off between computational load and communication efficiency. Similarly, HDC’s hypervector binding and summation operations, though computationally lightweight, demand extensive memory for high-dimensional storage. Existing resource management approaches are often static and fail to dynamically orchestrate resources across computation, communication, and representation layers simultaneously. This issue becomes pronounced across the device-edge-cloud platforms, characterized by varying degrees of resource availability and computational capabilities. Addressing these challenges requires the development of sophisticated adaptive optimization algorithms capable of dynamically adjusting to real-time network conditions, possibly leveraging real-time channel state information to adjust model splitting points, hypervector dimensionality, and transmission parameters. 
\subsection{Interpretability and Robustness in Dynamic Environments}
The opaque nature of large AI models and the unexplored robustness properties of HDC models under dynamic wireless conditions significantly impede their trustworthiness and deployment in critical scenarios. Large models used in intelligent transceivers, such as networks for beamforming or channel estimation, generally operate as closed boxes, making diagnosis of performance degradation caused by fluctuating channel conditions challenging. Similarly, while HDC offers inherent noise robustness, its behavior under dynamic environmental conditions such as Doppler effects, fading, and multipath remains insufficiently characterized. For example, small perturbations in hypervectors due to channel errors could accumulate during aggregation, leading to classification errors. Addressing these issues requires integrating interpretability and robustness techniques explicitly within ILAC frameworks. Explainable AI approaches such as attention visualization could elucidate the decision-making processes and highlight influential features or dimensions in neural networks and hypervectors, respectively. Research into adversarial robustness, sensitivity analyses, and anomaly detection methods tailored for dynamic communication environments is necessary to ensure reliable, interpretable, and resilient ILAC operations.

\subsection{Standardization, Protocal, and Testbed Development} The lack of common standards in both AI software ecosystems and wireless communication infrastructures poses a major obstacle for practical ILAC deployment. Popular AI frameworks such as TensorFlow and PyTorch, and communication standards like 5G NR and Wi-Fi 6, operate in isolation, making end-to-end optimization of learning and communication processes difficult. HDC’s hardware-software integration, such as FPGA-based hypervector accelerators, lacks industry-wide standards, leading to interoperability issues across different vendors.  This fragmentation leads to incompatibility in data formats, interface definitions, execution semantics, and performance metrics, resulting in isolated optimization efforts that cannot be reused or scaled efficiently across platforms. From a prototyping perspective, current hardware-software co-design tools for AI and communication are typically built with different assumptions about latency, determinism, and resource allocation. Integrating these heterogeneous tools to emulate ILAC performance under realistic scenarios is technically complex and labor-intensive. For example, deploying HDC algorithms on FPGA-based platforms alongside neural models requires consistent runtime environments and synchronized data pipelines, which are not yet standardized. 

To address these challenges, future research must develop open, cross-layer ILAC reference architectures that unify model deployment, network configuration, and runtime adaptation. This includes standardized APIs for model exchange, shared ontologies for system profiling, and integrated DevOps pipelines that support co-simulation of communication and learning tasks. Open-source initiatives, industry alliances, and regulatory support will be key drivers for establishing ILAC as a formalized computing and communication paradigm. Furthermore, establishing unified, accessible testbeds and prototypes, integrated into open frameworks like Open Radio Access Networks (Open RAN), is essential. These testbeds would provide a common experimental environment enabling rigorous validation and iterative refinement of ILAC methods. Efforts should also focus on developing standardized hardware-software interfaces, particularly for specialized technologies like FPGA-based hypervector accelerators, enhancing interoperability and scalability across diverse platforms and industries.
\subsection{Intelligent Agent Integration and Emerging Interaction Paradigms}
The integration of intelligent agents, such as autonomous robots, drones, self-driving vehicles, and other embodied systems, introduces unique challenges due to the inherent nature of intelligent agents, which are designed to interact with dynamic environments, make decisions in real-time, and perform tasks autonomously. These intelligent agents are supposed to constantly adapt their behavior and decisions based on sensory inputs, environmental feedback, and interactions with other agents. For instance, autonomous vehicle and drone systems need to continuously process sensor data (such as LIDAR, cameras, radar, etc.) to perceive their environment and make navigation decisions. The data processing and decision-making must occur in real-time, as any delay could lead to safety risks or failure to achieve the desired objectives. These systems also need to communicate with other agents (e.g., other vehicles, infrastructure) to share information, synchronize actions, and collaboratively achieve tasks.

The ILAC framework uniquely addresses these challenges by enabling joint optimization of both communication and learning processes. It provides a unified approach that allows intelligent agents to dynamically allocate resources between computation (for decision-making) and communication (for real-time data exchange with other agents and systems). This integrated approach ensures that the timely processing of sensory inputs and the efficient exchange of information are both optimized to maintain real-time responsiveness, particularly in complex and unpredictable environments. One potential method is adaptive learning-based resource allocation, allowing  intelligent agents to dynamically learn optimal resource allocation strategies based on environmental feedback and task importance. Additionally, edge computing plays a crucial role in reducing latency by processing data closer to the intelligent agent rather than relying solely on remote cloud servers.

\subsection{Scaling Laws and Performance Limits at Inference Level}
In the context of ILAC systems, understanding how performance scales with increasing model size, data complexity, and network resources is essential. Scaling laws, especially at the inference level, are not solely concerned with raw computational power but also how communication resources interact with computational demands. The ability to sustain high performance as these variables evolve introduces unique challenges, especially when considering the interplay between learning and communication processes. 

As AI models scale up, inference tasks become increasingly computationally demanding. Modern deep learning models, which can contain billions or even trillions of parameters, require substantial computational resources for real-time inference. In distributed environments, where inference is handled across multiple nodes, the communication overhead—particularly between edge nodes and the cloud—can quickly overwhelm available bandwidth. This results in bottlenecks that limit the scalability of large models, hindering real-time performance in applications that demand speed and efficiency. Therefore, achieving effective scaling of large models while ensuring timely and accurate inference remains a fundamental challenge, particularly in dynamic and complex real-world scenarios.

While cloud computing provides powerful computational resources, relying solely on the cloud for inference in resource-constrained environments, such as autonomous vehicles or mobile devices, creates significant latency and bandwidth issues. In these contexts, optimizing resource allocation between cloud and edge computing becomes imperative. As both the size of AI models and the complexity of data increase, it is crucial to effectively distribute computational, storage, and communication resources across the system. Failing to do so risks compromising inference performance, particularly in time-sensitive applications.

Moreover, in multi-agent environments, where multiple systems or agents must work in concert—such as in collaborative robotics or autonomous fleets of vehicles—scaling the system’s performance becomes even more complex. These systems require high throughput for data exchange between agents while maintaining low latency to enable real-time decision-making. As the number of nodes in the system increases, the demands on communication infrastructure also grow. Effectively managing both communication and computational resources across a large number of agents without sacrificing system performance is a critical challenge that must be addressed for large-scale deployments.

The ILAC framework represents a pivotal step toward intelligent 6G networks, but its success depends on resolving challenges in resource management, semantic integration, robustness, and standardization. By addressing these open problems through interdisciplinary research and collaborative standard-setting, ILAC can evolve into a scalable, trustworthy, and efficient framework that enables seamless integration of learning and communication, driving innovation in autonomous systems, smart cities, and beyond. 


\bibliographystyle{IEEEtran}
\bibliography{IEEEabrv,new_MMM}

\end{document}